\documentclass[reprint,superscriptaddress,amsmath,amssymb,amsthm,aps,prx,nofootinbib]{revtex4-2}

\usepackage{graphicx}
\usepackage{bm}
\usepackage{enumitem}
\usepackage[version=4]{mhchem}
\usepackage{inconsolata}
\usepackage[ruled,vlined,longend,algo2e]{algorithm2e}
\usepackage{titletoc}
\usepackage{setspace}

\usepackage[table,dvipsnames]{xcolor}
\definecolor{linkcolor}{HTML}{b91c1c}
\definecolor{citecolor}{HTML}{64748b}
\definecolor{urlcolor}{HTML}{2563eb}
\definecolor{liflow}{HTML}{ede9fe}

\usepackage{booktabs}
\usepackage{tabularray}
\UseTblrLibrary{booktabs}
\usepackage{makecell}
\usepackage{multirow}

\usepackage{hyperref}
\hypersetup{colorlinks,linkcolor=linkcolor,citecolor=citecolor,urlcolor=urlcolor}
\bibpunct{\color{citecolor}[}{\color{citecolor}]}{,}{n}{}{;}  
\usepackage[capitalise]{cleveref}

\usepackage{amsthm}
\newtheorem{proposition}{Proposition}

\renewcommand{\figurename}{Fig.}
\renewcommand{\tablename}{Table}
\renewcommand{\thetable}{\arabic{table}}

\crefname{extendedfigure}{Extended Data Fig.}{Extended Data Figs.}
\crefname{extendedtable}{Extended Data Table}{Extended Data Tables}

\crefname{suppfigure}{Fig.}{Figs.}
\crefname{supptable}{Table}{Tables}
\crefname{suppsection}{Supplementary Note Section}{Supplementary Note Sections}

\begin{document}

\title{Flow Matching for Accelerated Simulation of Atomic Transport \\ in Crystalline Materials}

\author{Juno Nam}
\affiliation{Department of Materials Science and Engineering, Massachusetts Institute of Technology, Cambridge, MA 02139, USA}
\affiliation{Energy Storage Research Alliance, Argonne National Laboratory, 9700 South Cass Avenue, Lemont, Illinois 60439, United States}
\author{Sulin Liu}
\affiliation{Department of Materials Science and Engineering, Massachusetts Institute of Technology, Cambridge, MA 02139, USA}
\author{Gavin Winter}
\affiliation{Department of Materials Science and Engineering, Massachusetts Institute of Technology, Cambridge, MA 02139, USA}
\author{KyuJung Jun}
\affiliation{Department of Materials Science and Engineering, Massachusetts Institute of Technology, Cambridge, MA 02139, USA}
\affiliation{Energy Storage Research Alliance, Argonne National Laboratory, 9700 South Cass Avenue, Lemont, Illinois 60439, United States}
\author{Soojung Yang}
\affiliation{Computational and Systems Biology Program, Massachusetts Institute of Technology, Cambridge, MA 02139, USA}
\author{Rafael G\'omez-Bombarelli}
\email{rafagb@mit.edu}
\affiliation{Department of Materials Science and Engineering, Massachusetts Institute of Technology, Cambridge, MA 02139, USA}
\affiliation{Energy Storage Research Alliance, Argonne National Laboratory, 9700 South Cass Avenue, Lemont, Illinois 60439, United States}

\date{\today}

\begin{abstract}
Atomic transport underpins the performance of materials in technologies such as energy storage and electronics, yet its simulation remains computationally demanding.
In particular, modeling ionic diffusion in solid-state electrolytes (SSEs) requires methods that can overcome the scale limitations of traditional \textit{ab initio} molecular dynamics (AIMD).
We introduce LiFlow, a generative framework to accelerate MD simulations for crystalline materials that formulates the task as conditional generation of atomic displacements.
The model uses flow matching, with a \textit{Propagator} submodel to generate atomic displacements and a \textit{Corrector} to locally correct unphysical geometries, and incorporates an adaptive prior based on the Maxwell--Boltzmann distribution to account for chemical and thermal conditions.
We benchmark LiFlow on a dataset comprising 25-ps trajectories of lithium diffusion across 4,186 SSE candidates at four temperatures.
The model obtains a consistent Spearman rank correlation of 0.7--0.8 for lithium mean squared displacement (MSD) predictions on unseen compositions.
Furthermore, LiFlow generalizes from short training trajectories to larger supercells and longer simulations while maintaining high accuracy.
With speed-ups of up to 600,000$\times$ compared to first-principles methods, LiFlow enables scalable simulations at significantly larger length and time scales.
\end{abstract}

\maketitle


\section{Introduction}
\label{sec:introduction}

Atomic transport is a fundamental process that governs the performance of materials in various technologies, including energy storage, catalysis, and electronic devices \citep{balluffi2005kinetics,yip2023molecular}.
Solid-state electrolytes (SSEs) are a prime example, emerging as a safer and more stable alternative to liquid electrolytes commonly used in lithium-ion batteries \citep{bachman2016inorganic}.
The study and design of SSEs rely on fast and accurate atomistic simulation techniques to model the intricate ionic diffusion behaviors that dictate the atomic transport in these materials.
The standard method, \textit{ab initio} molecular dynamics (AIMD), involves costly density functional theory (DFT) calculations for each propagation step in the scale of femtoseconds.
Hence, their application is limited to small spatiotemporal scales, often insufficient for characterizing diffusive dynamics or screening candidate materials.

Due to the high computational cost of \textit{ab initio} calculations, machine learning interatomic potentials (MLIPs) based on graph neural networks have been developed to approximate the results of the quantum calculations \citep{friederich2021machine,ko2023recent}.
Recent advances in universal MLIPs, such as MACE-MP-0 \citep{batatia2024foundation} and CHGNet \citep{deng2023chgnet}, enable faster simulations and linear scaling with respect to number of atoms, with optional fine-tuning to mitigate the softening effect caused by training frames from optimization trajectories in materials databases \citep{deng2024overcoming}.
However, even with MLIPs, dynamics must be discretized in sufficiently small time steps to ensure stable and accurate propagation \citep{fu2023forces} and are still too slow to enable scalable simulation to perform MD-based high-throughput screening from large material databases.

To accelerate MD simulations for small bio/organic molecules, several works have explored machine learning (ML) surrogates for time-coarsened dynamics by learning transition probability densities.
Timewarp (\citet{klein2024timewarp}) employs a conditional normalizing flow (CNF) with Markov chain Monte Carlo sampling, while Implicit Transfer Operator Learning (ITO, \citet{schreiner2024implicit}) is a conditional diffusion model designed as an arbitrary time-lag propagator.
Score Dynamics (SD, \citet{hsu2024score}) learns the score function of the transition density, F$^3$low (\citet{li2024f3low}) models protein CG frame transitions with flow matching, and Force-Guided Bridge Matching (FBM, \citet{yu2024force}) uses a conditional bridge process with a correction mechanism based on intermediate force fields.
Additionally, \citet{arts2023two} models coarse-grained (CG) dynamics with diffusion models.
These methods are applied to biomolecular simulations, with less chemical diversity and different symmetry requirements and task formulations from this work.
Notably, \citet{fu2023simulate} targets non-Markovian dynamics in CG polymer materials by learning the acceleration and using a score-based corrector.
While CG modeling allows for the explicit modeling of dynamics over longer timesteps using the equations of motion, modeling atomic transport in materials requires all-atom modeling, necessitating a generative surrogate for the dynamics.

This work aims to address this by developing a generative acceleration framework designed for scalable and cost-effective simulations of diffusive dynamics in crystalline materials across different temperatures.
The key objective is to construct a model capable of accurately reproducing relevant kinetic observables, such as mean squared displacement (MSD) and self-diffusivity of mobile ions, in comparison to long MD simulations using MLIPs or AIMD.
We formulate the task of conditional generation of atomic displacements, and we develop a flow matching approach with a physically-motivated adaptive prior to account for chemical and thermal conditions, along with a corrector mechanism to ensure stability.
Our approach accounts for periodic boundary conditions and generalizes effectively across different supercell sizes.


\section{Results}
\label{sec:results}

\begin{figure*}[!ht]
\includegraphics[width=0.8\textwidth]{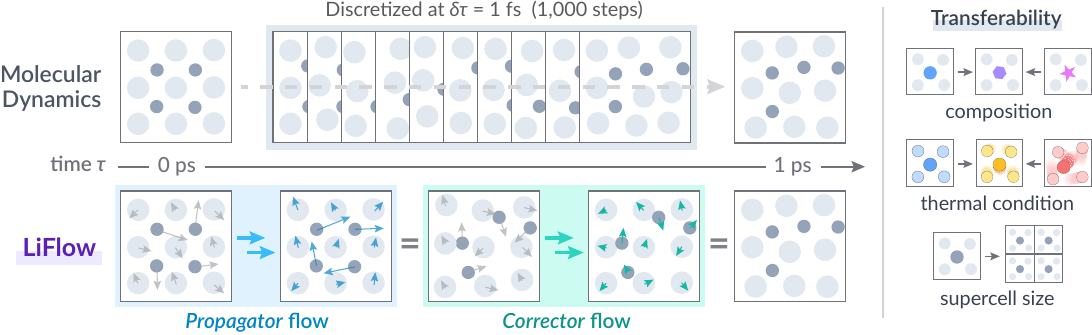}
\caption{
\textbf{LiFlow scheme.}
LiFlow is a generative acceleration framework for MD simulations for crystalline materials, with \textit{Propagator} and \textit{Corrector} components leveraging a conditional flow matching scheme for accurate generation of atomic displacements during time propagation.
Transferability across chemical composition, temperatures, and supercell sizes is considered in designing the task, adaptive prior distribution, and flow model architectures.
}
\label{fig:overview}
\end{figure*}

The LiFlow framework is illustrated in \cref{fig:overview}.
We begin by outlining the problem of generating atomic displacements to accelerate simulations and discussing the symmetry constraints necessary for scalable generation.
Next, we propose a physically-motivated prior and a flow model parametrization that adheres to these constraints, followed by the training and inference processes.

\subsection{Problem Setting}
\label{sec:problem_setting}

\textbf{Crystalline materials and representation.}
A crystal structure, assuming perfect order with translational symmetry, can be idealized as an infinite repetition of atoms, each assigned an atom type from the periodic table $\mathcal{A}$, within a unit cell with periodic boundary conditions \citep{ashcroft1976solid}.
In this work, a structure of a material with $n$ atoms in the unit cell is represented by the tuple $\mathcal{M} = (\bm{X}, \bm{L}, \bm{a})$, where $\bm{X} = (\bm{x}_1, \bm{x}_2, \cdots, \bm{x}_n)^\top \in \mathbb{R}^{n \times 3}$ denotes the Cartesian coordinates of the atoms, $\bm{L} = (\bm{l}_1, \bm{l}_2, \bm{l}_3)^\top \in \mathbb{R}^{3 \times 3}$ with rows defining the basis vectors of a 3-D repeating unit cell, and $\bm{a} \in \mathcal{A}^n$ is the atom types.
We impose a graph structure on the material by connecting pairs of nearby atoms with edges \citep{schutt2017schnet}, possibly across unit cell boundaries.
An edge $((i, j), \bm{k}) \in [n]^2 \times \mathbb{Z}^3$ is formed between atoms $i$ and $j$ if the distance between atom $i$ and atom $j$, displaced by $\bm{k}$ unit cells from $i$, is smaller than the cutoff, i.e., $\left\Vert \bm{x}_j + \bm{kL} - \bm{x}_i \right\Vert_2 < r_\text{cutoff}$.
An $a \times b \times c$ supercell of $\mathcal{M}$ is defined as
\begin{equation}
    (\bm{X}', \bm{L}', \bm{a}') = (\oplus_{\kappa=1}^{abc} (\bm{X} + \bm{1}_n \otimes \bm{k}_\kappa), \bm{L} \operatorname{diag}(a, b, c), \oplus_{\kappa=1}^{abc} \bm{a}),
    \label{eq:supercell}
\end{equation}
where $\oplus$ denotes concatenation, $\otimes$ is the outer product, and $\bm{k}_\kappa \in \mathbb{Z}_a \times \mathbb{Z}_b \times \mathbb{Z}_c$ represents the index of unit cell repetitions.

\textbf{Task setup.}
In this work, we denote physical time by $\tau$ and flow matching time by $t$.
Similar to previous ML-based MD acceleration methods in \cref{sec:introduction}, our goal is to model the transition probability density of a material structure over a time interval $\Delta \tau$, conditioned on the temperature $T$: $p(\mathcal{M}_{\tau + \Delta \tau} \vert \mathcal{M}_{\tau}, T)$.
For this task, we fix the lattice $\bm{L}$ (constant volume) and atom types $\bm{a}$, and set $\Delta \tau$ as 1 ps, which is 1,000 times larger than the usual MD time step of 1 fs \citep{marx2009ab}.
In MD simulations used to model the kinetics of materials, \textit{unwrapped} coordinates are utilized, meaning atomic coordinates are not confined to the unit cell, in order to keep track of the atomic displacements \citep{vonbulow2020systematic}.
As a result, unlike previous ML surrogates for dynamics of bio/organic molecules with a single connected component with the fixed center of mass, the distribution of positions does not have a finite support.
Therefore, we opt to model the distribution of \textit{displacements} over time interval $\Delta \tau$, $\bm{D}_{\Delta \tau} := \bm{X}_{\tau + \Delta \tau} - \bm{X}_{\tau}$.
In summary, the task is to learn the conditional distribution of atomic displacements $p(\bm{D}_{\Delta \tau} \vert \bm{X}_{\tau}, \bm{L}, \bm{a}, T)$ from a dataset of time-separated pairs of structures $\mathcal{D} = \left\{ ((\bm{X}_{\tau}, \bm{X}_{\tau + \Delta \tau}), \bm{L}, \bm{a}, T)) \right\}$, extracted from MD trajectories across various material compositions and temperatures.
More details and rationale on the task design choices can be found in \cref{sec:task_design}.

\subsection{Flow Matching for Time Propagation}
\label{sec:flow_matching_time_propagation}

\textbf{Flow matching.}
Flow matching \citep{lipman2023flow} is a generative modeling framework in which samples from the prior distribution $x_0 \sim p_0(x)$ are transported to samples from the data distribution $x_1 \sim q(x)$ by a time-dependent vector field $u_t(x)$ ($t \in [0, 1])$.
The vector field generates a flow $\psi_t$ defined with $\psi_0(x) = x$ and $(\mathrm{d}/\mathrm{d}t) \psi_t(x) = u_t(\psi_t(x))$ and a probability path $p_t(x) = [\psi_t]_\ast p_0(x)$.
The data conditional vector field $u_t(x \vert x_1)$ is available in closed form for the commonly used Gaussian probability path $p_t(x \vert x_1) = \mathcal{N}(x; \mu_t(x_1), \sigma_t(x_1)^2 \bm{I})$.
The marginal vector field model $v_t(x; \theta)$ is parametrized by a neural network and learned by the following conditional flow matching (CFM) regression objective:
\begin{align}
    & \mathcal{L}_\text{CFM}(\theta) = \mathbb{E}_{t \sim \mathcal{U}(t; 0, 1), x_1 \sim p_1(x), x \sim p_t(x \vert x_1)} \nonumber \\
    & \hspace{10em} \left\Vert v_t(x; \theta) - u_t(x \vert x_1) \right\Vert^2.
    \label{eq:cfm_loss}
\end{align}

\textbf{Symmetry considerations.}
The conditional probability density of displacements is invariant to permutation of atomic indices, global translation and lattice shift of atomic coordinates, global rotation applied to relevant variables, and supercell choice (we omit the physical time $\tau$ and $\Delta \tau$ here for brevity):
\begin{widetext}
\begin{align}
    & p(\bm{D} \vert \bm{X}, \bm{L}, \bm{a}, T) = p(\bm{P}\bm{D} \vert \bm{P}\bm{X}, \bm{L}, \bm{Pa}, T), & \bm{P} \in S_n \; \text{(permutation)} \label{eq:sym_perm} \\
    & p(\bm{D} \vert \bm{X}, \bm{L}, \bm{a}, T) = p(\bm{D} \vert \bm{X} + \bm{1}_n \otimes \bm{t}, \bm{L}, \bm{a}, T), & \bm{t} \in \mathbb{R}^3 \; \text{(global translation)} \label{eq:sym_global_trans} \\
    & p(\bm{D} \vert \bm{X}, \bm{L}, \bm{a}, T) = p(\bm{D} \vert \bm{X} + \bm{ZL}, \bm{L}, \bm{a}, T), & \bm{Z} \in \mathbb{Z}^{n \times 3} \; \text{(lattice periodicity)} \label{eq:sym_lattice} \\
    & p(\bm{D} \vert \bm{X}, \bm{L}, \bm{a}, T) = p(\bm{D} \bm{R} \vert \bm{X} \bm{R}, \bm{LR}, \bm{a}, T), & \bm{R} \in \mathrm{O}(3) \; \text{(rotation/reflection)} \label{eq:sym_rot} \\
    & p(\bm{D} \vert \bm{X}, \bm{L}, \bm{a}, T) = p(\bm{D}' \vert \bm{X}', \bm{L}', \bm{a}', T), & \; \text{(supercell, defined as \cref{eq:supercell})} \label{eq:sym_super}
\end{align}
\end{widetext}
In general, to model the invariant densities with CNFs, we need an invariant base distribution and equivariant flow vector fields \citep{kohler2020equivariant,klein2023equivariant}.
Translational invariances \cref{eq:sym_global_trans,eq:sym_lattice} and supercell invariance \cref{eq:sym_super} are satisfied by our choice of representation for materials.
For $\mathrm{O}(3)$ and $S_n$ symmetries, we model our prior and flow according to the following proposition.

\begin{proposition}
\label{prop:symmetry}
Given an invariant base distribution $p_0(\bm{D}_0)$ satisfying \cref{eq:sym_perm,eq:sym_rot} and an equivariant conditional vector field $u_t(\bm{D}_t \vert \bm{D}_1)$ with the following properties:
\begin{widetext}
\begin{align}
    u_t(\bm{P}\bm{D}_t \vert \bm{P}\bm{D}_1, \bm{P}\bm{X}, \bm{L}, \bm{P}\bm{a}, T) &= \bm{P} u_t(\bm{D}_t \vert \bm{D}_1, \bm{X}, \bm{L}, \bm{a}, T), &\qquad \bm{P} \in S_n \label{eq:flow_perm} \\
     u_t(\bm{D}_t\bm{R} \vert \bm{D}_1\bm{R}, \bm{X}\bm{R}, \bm{L}\bm{R}, \bm{a}, T) &= u_t(\bm{D}_t \vert \bm{D}_1, \bm{X}, \bm{L}, \bm{a}, T) \bm{R}, &\qquad \bm{R} \in \mathrm{O}(3) \label{eq:flow_rot}
\end{align}
\end{widetext}
the generated conditional probability path $p_{t \vert 1}(\bm{D}_t \vert \bm{D}_1)$ is invariant.
Furthermore, given that the data distribution $q(\bm{D}_1)$ is invariant, the marginal probability path $p_t(\bm{D}_t)$ is also invariant.
\end{proposition}

Note that the group actions of $S_n$ and $\mathrm{O}(3)$ on the optional conditional variable $\bm{D}_1$ are the same as their actions on $\bm{D}_t$.
The proof is given in \cref{si:proof_prop_1}.
We conducted an ablation study by replacing the PaiNN \citep{schutt2021equivariant} backbone in the LiFlow model with the GNS \citep{sanchezconzalez2020learning} backbone used in Orb \citep{neumann2024orb}, which is a state-of-the-art non-equivariant MLIP, to confirm that the generation task benefits from symmetry considerations (see \cref{si:equivariance_ablation}).

\textbf{Choice of prior distribution.}
While the normal distribution is commonly used in diffusion and flow-based generative models, incorporating task-specific inductive biases into the prior can improve the performance \citep{lee2022priorgrad,guan2023decompdiff,jing2023eigenfold,irwin2024efficient}.
The common goal in these methods is to reduce the transport cost by initializing the prior closer to the data distribution.
We use a physically-motivated prior based on the Maxwell--Boltzmann distribution, which additionally accounts for differences between atom types and reflects thermal and phase conditions.

We consider a Gaussian prior, $\bm{D}_0 \sim \mathcal{N}(\bm{D}_0; \bm{0}, \bm{\Sigma} \otimes \bm{I}_3)$, with a diagonal covariance $\bm{\Sigma} = \operatorname{diag}(\bm{\sigma})^2$, where $\bm{\sigma} = \bm{\sigma}(\bm{a}, T) \in \mathbb{R}^n$ is equivariant to atom index permutation.
This prior distribution satisfies the symmetry constraints \cref{eq:sym_perm,eq:sym_global_trans,eq:sym_lattice,eq:sym_rot,eq:sym_super}.
In MD simulation of materials, atomic displacements tend to be larger for lighter atoms and at higher temperatures.
In the short-time, non-interacting limit, the displacements can be expressed as $\bm{D}_{\delta \tau} = \dot{\bm{X}}_{\tau} \delta \tau$, where the marginal distribution of velocity follows the Maxwell--Boltzmann distribution, $\dot{\bm{X}}_{\tau} \sim \mathcal{N}(\dot{\bm{X}}_{\tau}; \bm{0}, \operatorname{diag}(k_\text{B} T / \bm{m}) \otimes \bm{I}_3)$, with $k_\text{B}$ being the Boltzmann constant.
Thus, it is reasonable to initialize the noise from a \textit{scaled} Maxwell--Boltzmann distribution with $\bm{\sigma} = \sigma \cdot (k_\text{B}T/\bm{m})^{1/2}$, where $\sigma$ is a constant hyperparameter controlling the scale.

In the specific context of simulations of kinetic processes in this work, where the simulations are often conducted at elevated temperatures, the material may undergo phase transitions (e.g., from solid to liquid) within the temperature range covered by the dataset.
Additionally, for lithium-based SSEs, lithium atoms may exhibit displacements several orders of magnitude larger than those of non-lithium (frame) atoms.
To account for these variations, we introduce a material-dependent \textit{adaptive} scaling factor for the Maxwell--Boltzmann distribution:
\begin{align}
    \bm{\sigma} &= \left[ \sigma_\text{Li}(\mathcal{M}_0, T) \cdot \mathbb{I}_{\bm{a} = \text{Li}} + \sigma_\text{frame}(\mathcal{M}_0, T) \cdot \mathbb{I}_{\bm{a} \neq \text{Li}} \right] \nonumber \\
    & \hspace{12em} \odot (k_\text{B}T/\bm{m})^{1/2},
    \label{eq:adaptive_prior}
\end{align}
where for each species $\mathcal{S} \in \{ \text{lithium}, \text{frame} \}$, $\sigma_\mathcal{S}$ selects a scale value from the hyperparameters $\{ \sigma_\mathcal{S}^\text{small}, \sigma_\mathcal{S}^\text{large} \}$ based on a binary classifier's prediction of whether the displacements for $\mathcal{S}$ will be small or large.
The classifier utilizes temperature and the average-pooled atomic invariant features of $\mathcal{S}$, extracted from a pre-trained MACE-MP-0 model \citep{batatia2024foundation}, based on the initial material structure $\mathcal{M}_0$.
Further details about the classifier model are provided in \cref{sec:prior_selector_model}.

\textbf{Flow parametrization.}
Following \citet{pooladian2023multisample}, we select the linear interpolation between the prior sample and the data sample as a conditional flow:
\begin{align}
    u_t(\bm{D}_t \vert \bm{D}_1) &= \frac{\bm{D}_1 - \bm{D}_t}{1 - t}, \\
    \bm{D}_t = \psi_t(\bm{D}_0 \vert \bm{D}_1) &= (1 - t) \bm{D}_0 + t \bm{D}_1.
    \label{eq:ot_flow}
\end{align}
This satisfies the symmetry constraints in \cref{eq:flow_perm,eq:flow_rot}.
The marginal flow approximator $v_t(\bm{D}_t, \bm{X}_\tau, \bm{L}, \bm{a}, T; \theta)$ should also respect these symmetry constraints.
We adopt the PaiNN model \citep{schutt2021equivariant} to balance expressiveness with inference speed.
The structure is encoded using a radial basis function expansion of atomic distances and the unit vector directions along edges (\cref{eq:painn_scalar_message,eq:painn_vector_message}).
We observed that encoding the intermediate structure $\bm{X}_\tau + \bm{D}_t$ significantly improves prediction performance (\cref{table:painn_ablation}).
Thus, we modify the message-passing layers of PaiNN to accept two structural inputs: $\bm{X}_\tau$ and $\bm{X}_\tau + \bm{D}_t$.
Further details on the model architecture and modifications are given in \cref{sec:flow_model_architecture}.

\textbf{\textit{Propagator} and \textit{Corrector} models.}
While, in theory, a single generative model should suffice to learn the density, prediction errors would arise from two sources: inaccuracies in the marginal flow prediction and discretization errors in the flow integration.
Moreover, since the trajectory generation is performed autoregressively, applying the generative model iteratively compounds these errors over time.
To address this, in addition to the flow matching model described earlier (\textit{Propagator}), we introduce an auxiliary flow matching model named \textit{Corrector}, inspired by \citet{fu2023simulate}, to rectify potential errors in the predicted displacements.

Although the \textit{Corrector} model is intended to correct errors in the final displacement resulting from the integration of \textit{Propagator}, directly mapping the generated output to an actual data sample to compute the target correction value can be complex, as it may require differentiating through the flow integration.
Therefore, we decouple the \textit{Propagator} and \textit{Corrector} models, training the \textit{Corrector} to denoise positional noise of arbitrary small scale.
Given a perturbed configuration $\tilde{\bm{X}}_\tau = \bm{X}_\tau + \bm{D}$, where the noise displacement is sampled from $\bm{D} \vert \bm{\sigma}' \sim \mathcal{N}(\bm{D}; \bm{0}, \operatorname{diag}(\bm{\sigma}')^2 \otimes \bm{I}_3$) with the noise scale $\bm{\sigma}' \sim \mathcal{U}(\bm{\sigma}'; \bm{0}, \sigma_\text{max} \bm{1}_n)$, the flow is trained to generate the possible denoising displacements $-\bm{D}$ conditioned on $\tilde{\bm{X}}_\tau$.

During inference, we alternate autoregressively between \textit{Propagator} and \textit{Corrector} flow integration for $N_\text{step}$ steps to generate a trajectory of length $N_\text{step} \Delta \tau$.
Further details on training and inference are provided in \cref{sec:training_and_inference}, with algorithms listed in \cref{si:training_and_inference_algorithms}.

\subsection{Universal MLIP Model}
\label{sec:universal_mlip_model}

\textbf{Universal MLIP dataset.}
To train a compositionally transferable generative model for time-shifting conformational distributions, long-time simulation trajectories that span a diverse range of compositional spaces in solid-state materials are required.
A total of 4,186 stable lithium-containing structures were retrieved from the Materials Project database \citep{jain2013commentary} to capture various modes of lithium-ion dynamics across different compositions.
For each structure, 25 ps MD simulations were performed using the MACE-MP-0 small universal MLIP model \citep{batatia2024foundation} at temperatures of 600, 800, 1000, and 1200 K, with a time step of 1 fs.
The dataset spans 77 elements across the periodic table, capturing a broad range of atomic environments and dynamic behaviors (\cref{fig:data}).
The dataset is divided into training (90\%) and test (10\%) sets based on material composition, with the validation set sampled from the training portion.
Details of the simulations and dataset statistics are provided in \cref{sec:datasets}.

\begin{table*}[!ht]
\caption{
\textbf{Results for the universal dataset.}
Evaluation metrics (\cref{sec:evaluation_metrics}) for various \textit{Propagator} priors (isotropic and uniform/adaptive scale Maxwell--Boltzmann) with or without the \textit{Corrector}.
\textit{Regressor}$^\dagger$ denotes a non-generative model predicting displacements directly.
\textit{P}[adaptive] + \textit{C} represents the baseline model without ablations.
Standard deviations (in parentheses) are from three independent runs.
MSD values are reported in units of \AA{}$^2$, and the best metric values for each experiment and temperature are shown in bold.
}
\footnotesize
\label{table:universal_result}
\begin{tblr}{
    colspec={cclccccc},rowsep=1pt,colsep=8pt,vline{4},
    cell{2}{1}={c=8}{halign=l},
    cell{3}{1}={r=3}{valign=m},
    cell{3}{2}={r=3}{valign=m},
    cell{6}{1}={c=8}{halign=l},
    cell{7}{1}={r=12}{valign=m},
    cell{7}{2}={r=3}{valign=m},
    cell{10}{2}={r=3}{valign=m},
    cell{13}{2}={r=3}{valign=m},
    cell{16}{2}={r=3}{valign=m},
}
\toprule
{\makecell{Train \\ $T$ (K)}} & {\makecell{Inference \\ $T$ (K)}} & {\makecell{Model}} & {\makecell{log MSD\textsubscript{Li} \\ MAE ($\downarrow$)}} & {\makecell{log MSD\textsubscript{Li} \\ $\rho$ ($\uparrow$)}} & {\makecell{log MSD\textsubscript{frame} \\ MAE ($\downarrow$)}} & {\makecell{RDF \\ MAE ($\downarrow$)}} & {\makecell{Stable traj. \\ \% ($\uparrow$)}} \\
\midrule
{\textbf{Exp 1.} Single temperature: Effect of Maxwell--Boltzmann prior} \\ \midrule
800 & 800  &  \textit{Regressor}$^\dagger$  & 1.636  & 0.535  & 0.876  & 0.416  & 90.2 \\
 &  &  \textit{P}[isotropic]  & 0.498 (0.003) & 0.753 (0.008)  & 0.318 (0.008) & 0.113 (0.0020) & 98.6 (0.2) \\
 &  &  \textit{P}[uniform]  & \textbf{0.396} (0.006) & \textbf{0.779} (0.009) & \textbf{0.274} (0.003) & \textbf{0.084} (0.0004) & \textbf{99.4} (0.1) \\
\midrule
{\textbf{Exp 2.} Multiple temperatures: Effects of adaptive prior scaling \cref{eq:adaptive_prior} and \textit{Corrector}} \\ \midrule
All  &  600  &  \textit{P}[uniform]  & \textbf{0.345} (0.003) & 0.740 (0.009) & 0.257 (0.006) & 0.082 (0.0001) & 99.8 (0.2) \\
 &  &  \textit{P}[adaptive]  & 0.376 (0.005) & 0.709 (0.003) & 0.286 (0.001) & 0.118 (0.0002) & 99.6 (0.2) \\
 &  &  \textit{P}[adaptive] + \textit{C}  & 0.348 (0.004) & \textbf{0.744} (0.012)  & \textbf{0.241} (0.002) & \textbf{0.069} (0.0001) & \textbf{100.0} (0.0) \\
\cmidrule{2-8}
 &  800  &  \textit{P}[uniform]  & 0.417 (0.007) & 0.737 (0.011) & 0.307 (0.003) & 0.091 (0.0005) & 99.8 (0.2) \\
 &  &  \textit{P}[adaptive]  & 0.385 (0.004) & 0.759 (0.008) & 0.294 (0.001) & 0.110 (0.0004) & 99.5 (0.0) \\
 &  &  \textit{P}[adaptive] + \textit{C}  & \textbf{0.366} (0.005) & \textbf{0.781} (0.005) & \textbf{0.255} (0.004) & \textbf{0.066} (0.0000) & \textbf{100.0} (0.0) \\
\cmidrule{2-8}
 &  1000  &  \textit{P}[uniform]  & 0.505 (0.011) & 0.705 (0.008) & 0.400 (0.007) & 0.124 (0.0006) & 98.6 (0.2)\\
 &  &  \textit{P}[adaptive]  & 0.456 (0.024) & 0.746 (0.008) & 0.374 (0.003) & 0.126 (0.0004) & 98.6 (0.6) \\
 &  &  \textit{P}[adaptive] + \textit{C}  & \textbf{0.429} (0.003) & \textbf{0.769} (0.006) & \textbf{0.332} (0.002)  & \textbf{0.071} (0.0001)  & \textbf{99.8} (0.1) \\
\cmidrule{2-8}
 &  1200  &  \textit{P}[uniform]  & 0.448 (0.006) & 0.788 (0.003) & 0.493 (0.003) & 0.168 (0.0013) & 95.5 (0.5) \\
 &  &  \textit{P}[adaptive]  & 0.410 (0.002) & 0.809 (0.003) & 0.416 (0.003) & 0.137 (0.0004) & 98.1 (0.6) \\
 &  &  \textit{P}[adaptive] + \textit{C}  & \textbf{0.389} (0.005)  & \textbf{0.821} (0.004) & \textbf{0.363} (0.003) & \textbf{0.079} (0.0002)  & \textbf{99.6} (0.1) \\
\bottomrule
\end{tblr}
\end{table*}

\begin{figure*}[!ht]
\includegraphics[width=\textwidth]{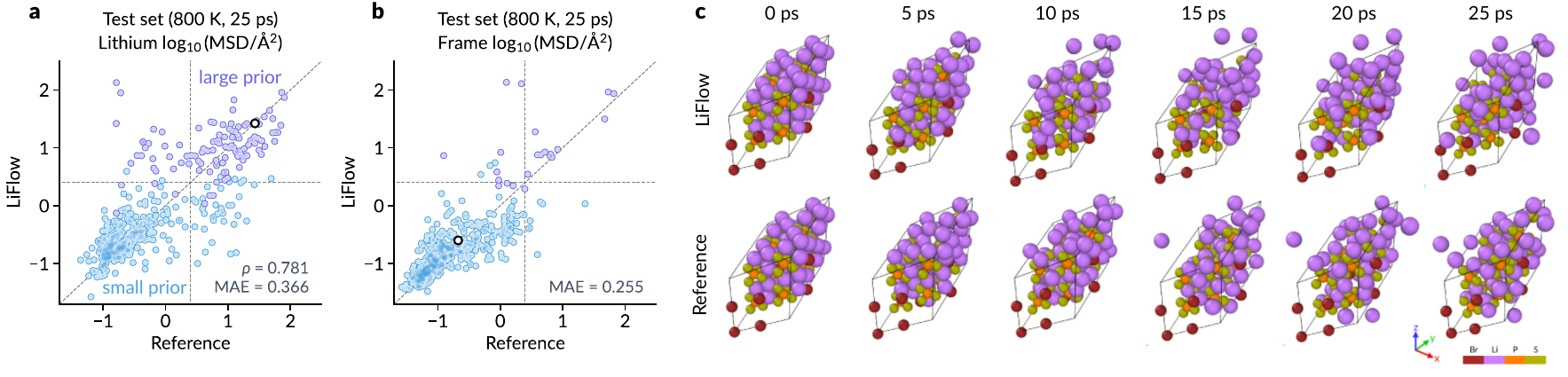}
\caption{
\textbf{Parity plots for kinetic metrics and trajectory visualizations.}
\textbf{(a, b)} Parity plots comparing the log MSD values for (a) lithium and (b) frame atoms in 800 K, 25 ps simulations across test set materials.
Data points are colored by their respective prior scales.
\textbf{(c)} Reference and generated trajectories for argyrodite \ce{Li6PS5Br} (highlighted points in a and b).
}
\label{fig:universal_result}
\end{figure*}

\textbf{Reproducing dynamic observables.}
We performed LiFlow inference iteratively for $N_\text{step} = 25$ steps to simulate dynamics over 25 ps.
We compared the log MSD values (\cref{eq:msd}) of lithium and frame atoms, assessing mean absolute error (MAE) and Spearman rank correlation between the reference and generated trajectories.
Additionally, we evaluated structural fidelity using radial distribution function (RDF) differences (\cref{eq:rdf}) and quantified the fraction of numerically stable generations.
Definitions and details of the evaluation metrics are provided in \cref{sec:evaluation_metrics}, and evaluation results on the test set are reported in \cref{table:universal_result}.

For LiFlow baseline model (\textit{P}[adaptive] + \textit{C}), we consistently observed a Spearman rank correlation of 0.7--0.8 for lithium MSD in unseen test compositions.
This indicates the potential of the LiFlow model for computational screening to identify materials with high lithium diffusivity.
The parity plot between log MSD values of reference and LiFlow-generated trajectories at 800 K, along with visualized example trajectories, is shown in \cref{fig:universal_result}.
We observed that the diffusive behavior in well-known SSEs in the test set is accurately reproduced, as shown in \cref{fig:universal_result}c for argyrodite \ce{Li6PS5Br}.
Trajectories with high error in metrics are visualized in \cref{fig:universal_example}.
Similar to challenges in classical or \textit{ab initio} MD with longer time steps, hydrogens often show fictitious diffusion due to their light mass, leading to large initial displacements under the Maxwell--Boltzmann distribution.
The \textit{Propagator} may struggle to match these with the smaller actual displacements.

Additionally, we investigated the possibility of batched inference for efficiency, sensitivity to hyperparameters, and the effects of dataset scaling in \cref{si:batched_inference,si:hyperparameter_sensitivity,si:dataset_scaling}.
Although trained primarily on lithium-containing materials, the model also qualitatively captures ion transport in sodium-based ionic conductors, as shown in \cref{si:ood_evaluation}.

\textbf{Effect of adaptive prior.}
First, in \cref{table:universal_result} (Exp 1), we compare the isotropic prior (\textit{P}[isotropic], $\bm{\sigma} = \sigma \cdot \bm{1}_n$), to the scaled Maxwell--Boltzmann prior (\textit{P}[uniform], $\bm{\sigma} = \sigma \cdot (k_\text{B}T/\bm{m})^{1/2}$), to evaluate the impact of atom-type-specific scaling on the prior.
To focus solely on the relative scale between atoms, we vary the scaling factor $\sigma$ for both the isotropic and Maxwell--Boltzmann priors at a fixed temperature (800 K), then compare the relevant metrics for the optimal $\sigma$ in each case.
The best results for isotropic ($\sigma = 10^{-1.5}$) and Maxwell--Boltzmann ($\sigma = 1$) priors are shown in the first row of \cref{table:universal_result}, and the results across all scales are provided in \cref{table:prior_iso_maxwell}.
The scaled Maxwell--Boltzmann prior outperforms the isotropic prior in reproducing all kinetic metrics (log MSD), confirming that the relative scaling of priors among elements is crucial for performance across a wide range of compositions.
Additionally, note that the poor performance of direct regression-based displacement prediction (\textit{Regressor}) highlights the necessity of generative modeling.

Next, in \cref{table:universal_result} (Exp 2), we apply the scale for \textit{P}[uniform] determined in the previous experiment to the training and inference on trajectories across all temperatures, and compare to the adaptive scale Maxwell--Boltzmann prior (\textit{P}[adaptive], \cref{eq:adaptive_prior}).
With the exception of the lowest temperature (600 K), where the prior classifier is mostly ineffective (see \cref{fig:prior_scale_hist}), the model using the adaptive prior outperforms the one with the uniform scale prior.
This suggests that the mixture-of-priors approach effectively guides the flow model in capturing the scale of atomic movements.

\textbf{Effect of \textit{Corrector}.}
In \cref{table:universal_result} (Exp 2), we then compare the \textit{Propagator}-only model (\textit{P}[adaptive]) and \textit{Propagator} + \textit{Corrector} model (\textit{P}[adaptive] + \textit{C}).
We observed improved reproduction of static structural features, indicated by lower RDF MAE, across all temperatures when using the \textit{Corrector} model.
Notably, all kinetic metrics also showed improvement with the use of \textit{Corrector}.
Since the \textit{Propagator} is a generative model of displacements conditioned on the current time step structure $\bm{X}_\tau$, correcting errors in the conditional structure improves the accuracy of the predicted cumulative displacements, as reflected in the MSD metric.

\subsection{AIMD Models}
\label{sec:aimd_models}

\textbf{AIMD datasets.}
To evaluate the ability to extend accurate atomistic dynamics from short AIMD simulations, we employed two sets of AIMD trajectories that exhibit diffusive lithium dynamics.
The first set includes \ce{Li3PS4} (LPS) simulations from \citet{jun2024nonexistence}, with $\sim$250 ps trajectories for 128-atom structures of $\alpha$-, $\beta$-, and $\gamma$-LPS, conducted at 600--800 K.
Among the three LPS polymorphs, $\alpha$- and $\beta$-LPS are fast lithium-ion conductors that remain stable at high temperatures, whereas the $\gamma$-phase is a slower lithium-ion conductor \citep{kimura2023stabilizing,lee2023weak}.
Their crystal structures are quite similar, primarily differing in the orientation of the \ce{PS4} tetrahedra and the corresponding lithium-ion sites, yet they exhibit drastically different lithium transport properties.

The second set comprises \ce{Li10GeP2S12} (LGPS, a prototypical lithium superionic conductor \citep{kamaya2011lithium}) simulations from \citet{lopez2024how}, which includes $\sim$150 ps MD trajectories for a $2 \times 2 \times 1$ supercell (200 atoms) of LGPS at temperatures of 650, 900, 1150, and 1400 K.
We used the first 25 ps of each trajectory as the training set.
See \cref{sec:datasets} for further details on dataset acquisition and processing.

\begin{figure*}[!ht]
\includegraphics[width=0.8\textwidth]{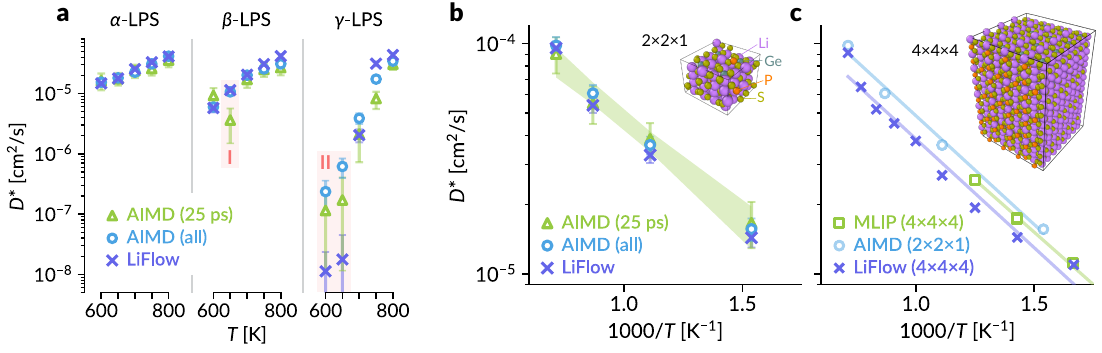}
\caption{
\textbf{Reproducing diffusivity from AIMD models.}
\textbf{(a)} Lithium self-diffusivity ($D^\ast$ for \ce{Li3PS4} (LPS) polymorphs, derived from AIMD (25 ps training, $\sim$250 ps full trajectories \citep{jun2024nonexistence}) and 250 ps LiFlow inference.
\textbf{(b)} Lithium $D^\ast$ as a function of $1000/T$ for \ce{Li10GeP2S12} (LGPS), using AIMD (25 ps training, $\sim$150 ps full trajectories \citep{lopez2024how}) and 150 ps LiFlow inference on a $2 \times 2 \times 1$ supercell.
Scatter points and error bars represent the median and 95\% confidence intervals (CIs) for $D^\ast$, from 1,000 MCMC samples of the Bayesian regression \citep{mccluskey2024kinisi}.
The shaded region represents the CIs for the Arrhenius fit ($1/T$ vs. $\log D^\ast(T)$) from 25 ps AIMD data.
\textbf{(c)} Results for a $4 \times 4 \times 4$ supercell from fine-tuned MLIP simulations \citep{winter2023simulations} and 1 ns LiFlow inference.
}
\label{fig:diffusion_AIMD}
\end{figure*}

\textbf{Reproducing kinetic observables.}
\cref{fig:diffusion_AIMD}a shows the reference lithium self-diffusivity values ($D^\ast$, \cref{eq:diffusivity}) for LPS from the AIMD simulations (25 ps for training and $\sim$250 ps full dataset) alongside the 250 ps LiFlow inference results.
Overall, the LiFlow results match the order of magnitude of the reference simulations, successfully reproducing the diffusivity differences among the LPS polymorphs.
This suggests that the model can detect subtle local structural variations between polymorphs and generate displacements accordingly.
In cases where diffusive behavior is expected but not sufficiently captured in a 25 ps trajectory to yield robust diffusivity statistics, LiFlow can \textit{infill} the correct diffusive dynamics based on other simulations (\cref{fig:diffusion_AIMD}a, box I).
Note that LiFlow is trained on a 25 ps trajectory but used for inference on the full 250 ps trajectory.
However, when lithium hopping events become exceedingly rare, as in $\gamma$-LPS at lower temperatures, the generative model suffers from mode collapse towards non-diffusive displacements, resulting in an underestimation of $D^\ast$ (\cref{fig:diffusion_AIMD}a, box II).

\begin{table}[!ht]
\caption{
\textbf{Activation energies for LGPS simulations.}
Activation energies and 95\% confidence intervals (CIs) derived from lithium diffusivity in AIMD simulations (25 ps, 250 ps) and LiFlow (250 ps), as shown in \cref{fig:diffusion_AIMD}b.
Note that the LiFlow model is trained on the 25 ps AIMD trajectory.
}
\label{table:activation_energy}
\begin{tblr}{colspec=cccc,rowsep=1pt,colsep=8pt,vline{3}} 
\toprule
Method & Time [ps] & $E_\text{A}$ [eV] & 95\% CI [eV] \\
\midrule
AIMD & 25 & 0.173 & (0.141, 0.205) \\
AIMD & 250 & 0.192 & (0.175, 0.205) \\
LiFlow & 250 & 0.185 & (0.181, 0.190) \\
\bottomrule
\end{tblr}
\end{table}

\cref{fig:diffusion_AIMD}b similarly presents the diffusivity values and their 95\% confidence intervals (CIs) for LGPS from AIMD simulations and LiFlow inference.
We also verify whether the temperature dependence of $D^\ast$ is accurately reproduced in terms of the activation energy $E_\text{A}$, a key measure of the lithium diffusion barrier.
For the $2 \times 2 \times 1$ supercell (\cref{fig:diffusion_AIMD}b), results in \cref{table:activation_energy} indicate that LiFlow value $E_\text{A} =$ 0.185 eV, is consistent with the reference AIMD value (0.192 eV).
Although the 25 ps AIMD $E_\text{A}$ (0.173 eV) lies outside the 95\% CI of the longer AIMD, LiFlow successfully matches the longer AIMD result and produces more reliable statistics with lower variance, thanks to its extended simulation rollouts.
Additionally, we confirmed that LiFlow trajectories capture correlated transport accurately, as reflected in the total diffusivity and Haven ratio reported in \cref{si:correlated_transport}.

\textbf{Large-scale inference.}
By modeling the distribution of atomic displacements, the generative model can naturally generalize across different supercell sizes, as indicated by the supercell invariance (\cref{eq:sym_super}).
We evaluated scalability and temperature transferability using a $4 \times 4 \times 4$ supercell, performing LiFlow inference over 1,000 steps (1 ns), with the resulting $D^\ast$ values presented in \cref{fig:diffusion_AIMD}c.
For temperatures below the maximum training temperature (1400 K), the LiFlow model generates stable trajectories that extend far beyond the 25 ps length of the training set trajectories.
When compared to the reference dynamics from \citet{winter2023simulations} on LGPS, which used extensive simulations with a fine-tuned MLIP, $D^\ast$ values predicted by LiFlow closely match the reference values within the training temperature range.
However, as we extend to much lower temperatures (higher $1000/T$) beyond the training range, $D^\ast$ decreases much more slowly than the reference values (\cref{fig:diffusion_LGPS_ext}), indicating fictitious diffusive behavior when extrapolating to lower temperatures.
This is expected, as the model was trained primarily on larger displacements at higher temperatures.

\textbf{Reproducing structural features.}
While reproducing kinetics is the main objective of this study, we additionally examined the reproduction of structural features, such as diffusion traces and probability densities of lithium positions.
The diffusion trace in \cref{fig:diffusion_trace} shows that the generated dynamics and the reference dynamics explore different but symmetrically related sites in unwrapped coordinates.
This confirms that the model is not merely memorizing the reference dynamics but is generalizing to physically equivalent configurations.
Additionally, the potentials of mean force (PMFs) for lithium atoms, shown in \cref{fig:pmf}, are accurately reproduced at lower temperatures and deviate slightly at higher temperatures, where they become noisier for LiFlow.
As displacements grow larger and more varied at higher temperatures, achieving high accuracy for static structural features becomes more challenging.

\begin{table}[!ht]
\caption{
\textbf{Prediction speed.}
Time required to predict a 1 ns trajectory for LGPS.
MLIP refers to the MACE-MP-0 small model \citep{batatia2024foundation}, and the AIMD simulation time is extrapolated from a shorter run.
Evaluation settings are detailed in \cref{sec:training_and_inference}.
}
\label{table:speed}
\begin{tblr}{colspec=ccrr,rowsep=1pt,colsep=8pt,vline{4},cell{4}{1}={r=2}{valign=m}} 
\toprule
Method & Supercell & \# atoms & Time \\
\midrule
AIMD & $2 \times 2 \times 1$ & 200 & 340 days \\
MLIP & $2 \times 2 \times 1$ & 200 & 5.8 hours \\
LiFlow & $2 \times 2 \times 1$ & 200 & 48 s \\
& $4 \times 4 \times 4$ & 3,200 & 352 s \\
\bottomrule
\end{tblr}
\end{table}

\textbf{Computational cost.}
The computation time for 1 ns of inference using the methods investigated in this paper is reported in \cref{table:speed}.
MLIP-based simulations significantly reduce the time required for materials simulations (days to hours), and the LiFlow model accelerates this even further (hours to seconds).
Even taking into account the training time of the LiFlow model ($\lesssim$ an hour), it remains significantly more efficient than AIMD simulations (see \cref{sec:training_and_inference}).
Given that AIMD scales as $\mathcal{O}(n^3)$ in theory, while both LiFlow and MLIPs scale as $\mathcal{O}(n)$ for large systems (assuming graphs with radius cutoffs), the LiFlow model enables efficient large-scale modeling of atomistic dynamics, as demonstrated in this work.


\section{Discussion}
\label{sec:discussion}

We proposed the LiFlow model, a generative acceleration framework designed to accelerate MD simulations for crystalline materials, with a focus on lithium SSEs.
The model consists of two key components: a \textit{Propagator}, which generates atomic displacements for time propagation, and a \textit{Corrector}, which applies denoising.
Both components utilize a conditional flow matching scheme, and we introduced a thermally and chemically adaptive prior based on the Maxwell--Boltzmann distribution and modified the PaiNN model as a marginal flow approximator, both of which were critical for the accurate reproduction of dynamics.
In our analysis of lithium-containing material trajectories, we consistently observed a Spearman rank correlation of 0.7--0.8 for lithium MSD in unseen compositions.
This indicates the potential of the LiFlow model for computational screening to identify materials with high lithium diffusivity.
Furthermore, we demonstrated the ability to extend short-length accurate AIMD trajectories by training the LiFlow model.
This allowed us to infill insufficient observations, reproduce accurate temperature dependencies, and maintain high accuracy when scaling up to much larger supercells.
Compared to simulations using MLIPs and AIMD, LiFlow offers significant speedups of 400$\times$ and 600,000$\times$, respectively.
This provides a practical means of scaling MD simulations to larger spatiotemporal domains.

There are several limitations and extensions remaining for future work.
First, although we have demonstrated the importance of designing the prior for the flow matching process, determining the appropriate prior scale remains a hyperparameter.
A theoretical analysis of the optimal prior distribution would provide a more principled approach to designing priors tailored to specific acceleration tasks and material systems.
This also applies to the choice of time step $\Delta \tau$: we used a fixed time step based on observation (\cref{sec:task_design}), but given the site-to-site hopping nature of atomistic transport, our method may benefit from adaptive or controllable time stepping (e.g., \citep{schreiner2024implicit}).
When combined with time acceleration via enhanced sampling methods \citep{tiwary2013metadynamics,bonati2021deep}, this could help alleviate mode collapse when hopping displacements are rare but important.
Additionally, while LiFlow performs well within the trained temperature range, it struggles to extrapolate beyond the training regime, where system dynamics may differ significantly from the training data.
As a result, the current approach lacks the broad generalizability seen in universal MLIP models, which preserve the physical dynamics of systems while approximating the potential energy landscape.
The prediction of low-temperature behavior relies on extrapolating observables from the training temperature range, rather than directly capturing the underlying dynamics at the target temperature.
Consequently, with the current methodology, the practically important regime of slow and strongly correlated ion transport at room temperature remains difficult to address.
To improve reliability and develop a model capable of capturing emergent system behaviors, generative propagators would benefit from incorporating thermodynamic design principles \citep{tiwary2024generative,dibak2022temperature,herron2023inferring}, apply inductive biases from the stationary distribution \citep{diez2025boltzmann}, and enforce detailed balance under equilibrium conditions \citep{mardt2020deep,klein2024timewarp}.
To extend to atomic transport in non-equilibrium steady states, such as ionic diffusion under an electric field, adapting the existing equivariant network framework to account for responses to external fields would be useful \citep{falletta2025unified}.
Lastly, the accuracy of LiFlow depends on the compositional coverage and accuracy of the reference dynamics.
In discovery settings, uncertainty estimates \citep{imbalzano2021uncertainty,tan2023single} are needed to determine when to collect additional reference trajectories for further training in unexplored compositional spaces.
Given the range of MD simulation methods and their trade-offs between accuracy and speed, transfer learning or multi-fidelity frameworks \citep{meng2020composite} may also be useful.


\section{Methods}
\label{sec:methods}

\subsection{Task Design}
\label{sec:task_design}

\textbf{Fixing the volume.}
In AIMD simulations for solid electrolytes, or in general when modeling the transport properties of atomistic systems, simulations are typically conducted under the NVT (constant volume) ensemble.
Although real materials are often under constant pressure conditions, employing a barostat in simulations to control pressure modifies cell volume, potentially leading to significant changes in particle positions and dynamics \citep{maginn2018best}.
In practice, AIMD simulations are initiated after energy minimization of the material structure (with respect to both atomic coordinates and cell dimensions) under the assumption that thermal expansion of the cell does not significantly affect the transport properties.

\textbf{Unwrapped coordinates.}
In atomistic systems with periodic boundary conditions (PBCs), particles that exit one side of the simulation box effectively reenter from the opposite side.
A straightforward way to handle this is to use \textit{wrapped} coordinates, where the positions are continuously confined within the simulation box.
However, this introduces jumps in atomic positions during long-range motions, which can distort the calculation of kinetic properties such as MSD and diffusivity.
To avoid this, the coordinates must be \textit{unwrapped} before computing such properties.
Alternatively, particle positions can be propagated using unwrapped coordinates from the start, without wrapping them back when crossing the cell boundaries.

It is possible to unwrap trajectories during the post-processing of AIMD simulations, assuming that no particles move more than half the cell dimensions between time steps.
This condition generally holds for typical AIMD simulations, which use small time steps.
However, in the case of LiFlow modeling in this work, particle displacements can exceed half the box size because (1) we simulate with a much larger time step $\Delta \tau$, and (2) AIMD simulation cells are typically small due to high computational costs.
Hence, we use unwrapped coordinates directly when formulating the displacement modeling task for LiFlow.

As a time-hopping conditional generative model for material structures, our approach shares design principles with crystal generation models \citep{xie2022crystal,jiao2023crystal,ai4science2023crystal,zeni2024matter,yang2024scalable,miller2024flowmm}, which use diffusion or flow matching to generate atomic identities and positions within a unit cell.
While these methods often handle position generation as fractional coordinates with periodic boundaries, our task requires modeling displacements in Cartesian coordinates directly without wrapping positions back into the unit cell.

\textbf{Choice of $\Delta \tau$.}
Since the goal of generative displacement modeling in this work is to efficiently accelerate MD simulations, the propagation time step $\Delta \tau$ must be significantly larger than the MD time step $\delta \tau$.
However, due to the high cost of generating data, $\Delta \tau$ should not be so large that the modes of atomic displacements are not adequately covered by the training set trajectories.

To determine $\Delta \tau$, we consider the time evolution of lithium MSD for typical lithium-ion SSEs.
For small $\Delta \tau$ values ($<$ 0.1 ps), the MSD grows approximately as $\text{MSD} \propto \Delta \tau^{1.42}$, reflecting the ballistic and vibrational motion of lithium ions \citep{he2018statistical}.
In this regime, the benefit of generative modeling is limited, as the evolution of atomic positions is closely related to the initial velocities.
For larger $\Delta \tau$ ($\gtrsim$ 1 ps), the MSD grows linearly as $\text{MSD} \propto \Delta \tau$, indicating the onset of diffusive motion, as described by \cref{eq:diffusivity}.
Given that our training trajectories span 25 ps, we select $\Delta \tau =$ 1 ps to ensure that the generative model captures a diverse range of displacement modes present in the training data.

\textbf{Units.}
The atomic unit system is adopted in this work.
Unless stated otherwise, the units are as follows: length is in \AA{}, temperature in K, mass in atomic mass units (u), and energy in eV.
For example, the scaling factor for the Maxwell--Boltzmann prior has an implied unit of $\text{\AA{}}\cdot (\text{eV} \cdot \text{K} / \text{u})^{-1/2}$ for converting $(k_\text{B}T/\bm{m})^{1/2}$ into positions.

\subsection{Datasets}
\label{sec:datasets}

\textbf{Universal MLIP dataset.}
We fetched 4,186 lithium-containing structures from Materials Project \citep{jain2013commentary} with the criteria of (1) more than 10\% of the atoms are lithium, (2) band gap $>$ 2 eV, and (3) energy over the convex hull $<$ 0.1 eV/atom.
These criteria are designed to sample various modes of lithium-ion dynamics across different compositions, while maintaining minimal requirements for the SSEs.
After building a supercell of the structure in order to ensure that each dimension is larger than 9 \AA{} and minimizing the structure, we conducted NVT MD simulations with MACE-MP-0 small model \citep{batatia2024foundation} at 600, 800, 1000, and 1200 K for each structure.
The initial velocities were assigned according to the temperature, and the system was propagated for 25 ps with the time step of 1 fs (25,000 steps) using Nos\'e--Hoover dynamics \citep{nose1984unified,hoover1985canonical} as implemented in ASE \citep{larsen2017atomic}.
We recorded the atom positions every ten steps.
Note that, to avoid thermostat-dependent dynamical artifacts, velocity \textit{scaling} thermostats (e.g., Berendsen, Nos\'e--Hoover, and stochastic velocity rescaling) should be used instead of velocity \textit{randomization} thermostats (e.g., Langevin and Andersen).
The latter may lead to reduced diffusivity values due to rapid decorrelation of velocities \citep{basconi2013effects}.
We used pymatviz package \citep{riebesell2022pymatviz} to create dataset statistics in \cref{fig:data}.

\textbf{AIMD datasets.}
We used the LPS trajectories from \citet{jun2024nonexistence}.
Supercell sizes of $2 \times 2 \times 2$, $1 \times 2 \times 2$, and $2 \times 2 \times 2$ were used for $\alpha$-, $\beta$-, and $\gamma$-\ce{Li3PS4}, respectively.
For each structure, five trajectories at temperatures of 600, 650, 700, 750, and 800 K were used.
The reference trajectories used a time step of $\delta \tau =$ 2 fs, which we subsampled every five steps to reduce redundancy in the training and test datasets.
We set the LiFlow time step $\Delta \tau$ to 500 steps (1 ps).

For LGPS, we utilized the trajectories from \citet{lopez2024how}.
The reference simulations employed a time step of $\delta \tau =$ 1.5 fs, with snapshots recorded every ten steps (15 fs).
To align with this, we set the LiFlow time step $\Delta \tau$ to 670 steps (1.005 ps).

\subsection{Prior Selector Model}
\label{sec:prior_selector_model}

The prior selector model $\sigma_\mathcal{S}(\mathcal{M}_0, T)$ for species $\mathcal{S}$ (lithium or frame) is a binary classifier that predicts whether the atom of the given species $\mathcal{S}$ will exhibit large or small displacements based on the initial structure of materials.
The same training and test splits were used for the universal dataset.
Labels for large and small displacements were determined by the criterion $\text{MSD}_\mathcal{S} / \tau < 0.1$ \AA{}$^2$/ps, computed over the reference simulation ($\tau = 25$ ps).
The input features for the classifier are the atomic invariant features (128 dimensions) averaged over atoms of $\mathcal{S}$, extracted from a pre-trained MACE-MP-0 small model \citep{batatia2024foundation} given the initial structure $(\bm{X}_0, \bm{L}, \bm{a})$, along with the temperature ($T / 1000$ K, a scalar).
These features are concatenated and fed into a multi-layer perceptron with hidden layers of size 32 and 16, which is trained on the training set materials.
The histograms of the $\text{MSD}_\mathcal{S}/\tau$ distribution annotated with predicted labels are reported in \cref{fig:prior_scale_hist}.

\subsection{Flow Model Architecture}
\label{sec:flow_model_architecture}

We adapt the PaiNN model \citep{schutt2021equivariant} to parametrize the marginal flow approximator $v_\theta(\bm{D}_t \vert \bm{X}_\tau, \bm{L}, \bm{a}, T)$ for both the \textit{Propagator} and \textit{Corrector}.
\citet{schreiner2024implicit} employed a modified version of the PaiNN model, named ChiroPaiNN, for a similar task for small biomolecules, introducing cross products during message passing in order to break reflection symmetry.
Their modification was necessary due to their use of coarse-grained protein representation (C$_\alpha$ coordinates), where the mirror image of a C$_\alpha$ trace does not correspond to the mirror image of the full-atom structure.
In contrast, we represent the material structure using all atomic coordinates without coarse-graining, preserving the reflection symmetry of the atomistic system.
As a result, we chose to modify the original PaiNN architecture instead of ChiroPaiNN.

\textbf{Node input features.}
The model employs a learnable atomic embedding function, $f_\text{atom}: \mathcal{A} \to \mathbb{R}^{d_f}$, to map atomic species to feature vectors, where $d_f$ is the feature dimension.
For continuous values, the embedding function $f_\text{cont}: \mathbb{R} \to \mathbb{R}^{d_f/2}$ is defined using a sinusoidal encoding:
\begin{equation}
    \left[f_{\text{cont}}(x)\right]_i = \begin{cases}
        \sin \left( 2 \pi f_{\lfloor i/2 \rfloor} x \right) & i \text{ odd}, \\
        \cos \left( 2 \pi f_{\lfloor i/2 \rfloor} x \right) & i \text{ even},
    \end{cases}
    \label{eq:emb_cont}
\end{equation}
where $f_i$ ($i \in [d_f/4]$) are frequencies sampled from a standard normal distribution $\mathcal{N}(0, 1^2)$ and fixed during training.
The invariant node embedding for atom $j$ is computed as $f_\text{atom}(a_j) + (f_\text{cont}(T / 1000) \oplus f_\text{cont}(t))$, for temperature $T$ and flow matching time $t$.
Rather than initializing equivariant node features to zeros as in the original model, they are initialized from the current step displacement as $\bm{D}_t \otimes \bm{w} \in \mathbb{R}^{n \times 3 \times d_f}$, where $\bm{w} \in \mathbb{R}^{d_f}$ is a learnable weight vector.

One major difference between previous temperature-conditioned generative models (e.g., \citet{dibak2022temperature}) and ours is that we apply temperature conditioning to both the prior and the flow model input, as described here.
This is because we model displacements to track atomic movements across periodic boundaries over large time intervals.
Atomic configurations at higher temperatures resemble those at lower temperatures, so the flows or scores generating these configurations are expected to be similar.
However, atomic displacements in crystalline materials scale with temperature according to an Arrhenius relationship, growing exponentially with $-1/T$.
Empirically, we found it beneficial to condition the prior to initially separate the displacement scale, and to train the flow model to match the distribution more effectively, rather than relying on a single component to do so.

\textbf{Message passing.}
For clarity and ease of comparison, we use the notation from the PaiNN paper \citep{schutt2021equivariant} for this part.
As described in the main text, we leverage information from two sets of coordinates, $\bm{X}_\tau$ and $\bm{X}_\tau + \bm{D}_t$, during message passing.
A similar approach using two sets of edge information was previously employed by \citet{hsu2024score}.
To simplify computation, we define the edges using a radius cutoff graph based on $\bm{X}_\tau$, avoiding the need to reconstruct the neighbor graph at each flow matching step $t$.
When expanding distances into radial basis functions, we shift the distance by 0.5 \AA{}.
Unlike physically realistic atomistic systems, during flow integration, the structure $\bm{X}_\tau + \bm{D}_t$ may experience atomic clashes.
Since Bessel function values change most significantly at small radii, shifting the distances helps reduce variance in the edge features.

In the message functions that use continuous-filter convolutions, we apply elementwise addition of the filters corresponding to the two distances, $\Vert \vec{r}_{ij, 1} \Vert$ and $\Vert \vec{r}_{ij, 2} \Vert$.
To avoid introducing unintended permutation symmetry between two geometries, we use two distinct filters ($\mathcal{W}_{vs,k}'$ in \cref{eq:painn_vector_message_mod}) for the respective unit vector directions.
The invariant message update \cref{eq:painn_scalar_message} (Eq.~(7) in the original paper) is modified as \cref{eq:painn_scalar_message_mod}:
\begin{align}
    \Delta \mathbf{s}_i^m &= \sum_j \bm{\phi}_s(\mathbf{s}_j) \circ \mathcal{W}_s(\Vert \vec{r}_{ij} \Vert), \label{eq:painn_scalar_message} \\
    \Delta \mathbf{s}_i^m &= \sum_j \bm{\phi}_s(\mathbf{s}_j) \circ \left[ \mathcal{W}_s(\Vert \vec{r}_{ij,1} \Vert) + \mathcal{W}_s(\Vert \vec{r}_{ij,2} \Vert) \right], \label{eq:painn_scalar_message_mod}
\end{align}
and the equivariant message update \cref{eq:painn_vector_message} (Eq.~(8)) in the original paper) is modified as \cref{eq:painn_vector_message_mod}:
\begin{align}
    \Delta \vec{\mathbf{v}}_i^m &= \sum_j \vec{\mathbf{v}}_j \circ \bm{\phi}_{vv}(\mathbf{s}_j) \circ \mathcal{W}_{vv}(\Vert \vec{r}_{ij} \Vert) \nonumber \\
    & \quad + \sum_j \bm{\phi}_{vs}(\mathbf{s}_j) \circ \mathcal{W}_{vs}'(\Vert \vec{r}_{ij} \Vert) \frac{\vec{r}_{ij}}{\Vert \vec{r}_{ij} \Vert}, \label{eq:painn_vector_message} \\
    \Delta \vec{\mathbf{v}}_i^m &= \sum_j \vec{\mathbf{v}}_j \circ \bm{\phi}_{vv}(\mathbf{s}_j) \circ \left[ \mathcal{W}_{vv}(\Vert \vec{r}_{ij,1} \Vert) + \mathcal{W}_{vv}(\Vert \vec{r}_{ij,2} \Vert) \right] \nonumber \\
    & \quad + \sum_{k \in \{1, 2\}} \sum_j \bm{\phi}_{vs,k}(\mathbf{s}_j) \nonumber \\
    & \qquad \circ \left[ \mathcal{W}_{vs,k}'(\Vert \vec{r}_{ij,1} \Vert) + \mathcal{W}_{vs,k}'(\Vert \vec{r}_{ij,2} \Vert) \right] \frac{\vec{r}_{ij,k}}{\Vert \vec{r}_{ij,k} \Vert}. \label{eq:painn_vector_message_mod}
\end{align}

\textbf{Performance comparison.}
Since we use $\bm{D}_t$ to initialize the vector node features, the additional positional input $\bm{X}_\tau + \bm{D}_t$ could be omitted without losing information, allowing the use of the original PaiNN model.
\cref{table:painn_ablation} presents a comparison of the metrics from \cref{table:universal_result} between the original and modified PaiNN models.
The results show a significant difference between the two models, highlighting the importance of incorporating the intermediate structure $\bm{X}_\tau + \bm{D}_t$.
Furthermore, in \cref{fig:ablation_equiv}, we compared the equivariant PaiNN backbone with the non-equivariant GNS backbone \citep{sanchezconzalez2020learning} with rotational augmentation, and found the equivariant architecture to be more effective for the current task.

\subsection{Training and Inference}
\label{sec:training_and_inference}

\textbf{LiFlow training.}
We train the model using time-separated pairs of structures, $((\bm{X}_{\tau}, \bm{X}_{\tau'}), \bm{L}, \bm{a}, T))$, sampled from MD trajectories in the training set.
First, the prior displacements are sampled based on the possible choices outlined in \cref{sec:flow_matching_time_propagation}.
The \textit{Propagator} and \textit{Corrector} are trained to approximate the marginal flows toward the distributions of the possible propagating displacements, $\bm{X}_{\tau'} - \bm{X}_\tau$, and denoising displacements, $\bm{X}_{\tau'} - \tilde{\bm{X}}_{\tau'}$, respectively.
These are conditioned on the previous structure, $\bm{X}_\tau$, and the noisy structure, $\tilde{\bm{X}}_{\tau'}$, respectively.
Given interpolated displacements $\bm{D}_t$ and the corresponding conditional variables, both models are trained to match the ground truth conditional flow $u_t(\bm{D}_t \vert \bm{D}_1)$ using the regression loss \cref{eq:cfm_loss}.
Detailed training algorithms are reported in \cref{si:training_and_inference_algorithms}.

\textbf{LiFlow inference.}
Starting from the initial atom positions $\bm{X}_0$, we alternate between \textit{Propagator} and \textit{Corrector} flow integration for $N_\text{step}$ steps, generating the trajectory $\{ \bm{X}_0, \bm{X}_{\Delta \tau}, \cdots, \bm{X}_{N_\text{step}\Delta \tau} \}$.
The flow integration for both the \textit{Propagator} and \textit{Corrector} begins by sampling prior displacements $\bm{D}_0$ from the chosen prior distribution.
These displacements are then updated over $N_\text{flow}$ steps using Euler's method, based on the predicted marginal flow.
Since MD simulations are often performed with a fixed center-of-mass (CoM) position, defined as $\text{CoM}(\bm{X}, \bm{m}) = \sum_j m_j \bm{x}_j / \sum_j m_j$, we correct for any CoM drift after each \textit{Propagator}--\textit{Corrector} inference step.

\textbf{Training and inference hyperparameters.}
The training and model hyperparameters are summarized in \cref{table:hyperparameter}.
Additionally, validation loss was evaluated every 1,250 training steps, with early stopping triggered if the validation loss did not improve after ten evaluations.
The model parameters corresponding to the lowest validation loss were used for inference.

For the adaptive prior in the universal dataset, we set the scale hyperparameters as $(\sigma_\text{Li}^\text{low}, \sigma_\text{Li}^\text{high}, \sigma_\text{frame}^\text{low}, \sigma_\text{frame}^\text{high}) = (1, 10, 10^{-0.5}, 10^{0.5})$, based on the observation that lithium atoms are generally more diffusive than the frame atoms.
For \textit{Corrector} model, we used a maximum noise scale of $\sigma_\text{max} = 0.25$ and a small uniform-scale Maxwell--Boltzmann prior with $\sigma = 0.1$.
We conducted LiFlow inference iteratively for $N_\text{step} = 25$ steps to simulate dynamics over 25 ps with a time step of $\Delta \tau = 1$ ps.
Each inference step involves $N_\text{flow} = 10$ flow matching iterations of both \textit{Propagator} and \textit{Corrector} models.
During each inference process for a given structure, we terminated when either the maximum number of steps ($N_\text{step}$) was reached or the model prediction diverged due to instabilities.

For \textit{Propagator} in the AIMD dataset, we replaced the prior classifier with fixed prior scale parameters for each temperature, determined based on the MSD values from the training trajectories.
Additionally, in both LGPS and LPS, the frame atoms did not exhibit diffusive behavior, so we applied a uniform prior scale for these atoms.
The prior scales used were $(\sigma_\text{Li}^\text{small}, \sigma_\text{Li}^\text{large}) = (1, 10)$ for lithium atoms, and $\sigma_\text{frame} = 0.5$ for LGPS and $1$ for LPS.
For the \textit{Corrector}, we set the maximum noise scale to $\sigma_\text{max} = 0.1$ for the $2 \times 2 \times 1$ supercell of LGPS and for all LPS experiments.
For the larger $4 \times 4 \times 4$ supercell inference in LGPS, we used a \textit{Corrector} trained with $\sigma_\text{max} = 0.2$ to improve trajectory stability.

We performed LiFlow inference for $N_\text{step} = 150$ steps in the $2 \times 2 \times 1$ LGPS simulations and $N_\text{step} = 1000$ steps in the $4 \times 4 \times 4$ LGPS simulations, with a time step of $\Delta \tau = 1.005$ ps.
This corresponds to total simulation times of 150.75 ps and 1.005 ns, respectively.
For the LPS simulations, we used $N_\text{step} = 250$ steps with a time step of $\Delta \tau = 1$ ps, resulting in a total simulation time of 250 ps.
We used Euler integration with $N_\text{flow} = 10$ steps for all experiments.
For AIMD simulations, since the \textit{Propagator} error is relatively small, \textit{Corrector} inference can be simplified without impacting simulation results---for example, by reducing $N_\text{flow}$ to $1$.
Details of these ablation studies are provided in \cref{si:hyperparameter_sensitivity}.

\textbf{Implementation details.}
We implemented the LiFlow model using PyTorch \citep{paszke2019pytorch} and PyG \citep{fey2019fast} libraries.
For MLIP-based simulations, we utilized MACE-MP-0 small model (\texttt{mace-torch} package \citep{batatia2024foundation}) in combination with ASE \citep{larsen2017atomic}.
For the universal MLIP dataset, 25-ps NVT MD for 4,186 material compositions at four different temperatures took approximately 400 hours over eight NVIDIA Tesla V100 GPUs without CUDA optimization for equivariant models.
Bayesian analysis of diffusivity and activation energy was performed using the \texttt{kinisi} package \citep{mccluskey2024kinisi}.
Training and inference of LiFlow models were performed using a single NVIDIA RTX A5000 GPU.
The training process for the \textit{Propagator} and \textit{Corrector} models, using early stopping, typically lasts between 45,000 and 70,000 steps.
This corresponds to approximately 40--60 minutes of training, extending to up to two hours if the maximum step budget is reached.
For AIMD simulation in \cref{table:speed}, we used the $\Gamma$-point only version of VASP (\texttt{vasp\_gam} \citep{hafner2008ab}) with 48 cores of an Intel Xeon Gold 8260 CPU.
The same input files used in the LGPS AIMD simulations were utilized for the benchmark.

\subsection{Evaluation Metrics}
\label{sec:evaluation_metrics}

To quantify the prediction of kinetic observables, we compared the MSD of lithium and frame atoms between generated and reference trajectories.
The MSD measures the average squared distance that particles of type $\mathcal{S}$ move over time $\tau$:
\begin{align}
    \text{MSD}_\mathcal{S}(\tau) &= \frac{1}{\vert \mathcal{S} \vert} \sum_{i \in \mathcal{S}} \Vert \bm{x}_{\tau, i} - \bm{x}_{0, i} \Vert^2.
    \label{eq:msd}
\end{align}
Given the wide range of magnitudes of MSD values, we compared the log values (base 10) of MSD, with MSD in units of \AA{}$^2$.
We report the mean absolute error (MAE) and Spearman's rank correlation ($\rho$) for the log MSD predictions on the universal MLIP dataset.

In the long-time limit, the MSD grows linearly with time, with a rate proportional to the self-diffusivity $D_\mathcal{S}^\ast$:
\begin{align}
    D_\mathcal{S}^\ast &= \lim_{\tau \to \infty} \frac{\text{MSD}_\mathcal{S}(\tau)}{6 \tau}.
    \label{eq:diffusivity}
\end{align}
This is quantified using Bayesian regression of MSD against time \citep{mccluskey2024accurate,mccluskey2024kinisi}.
We further calculate the activation energy $E_\text{A}$ from the temperature dependence of diffusivity using the Arrhenius relationship $\log D^\ast(T) = \log D_0^\ast - E_\text{A} / k_\text{B}T$.

To evaluate the reproduction of structural features, we compare the all-particle radial distribution function (RDF), $g(r)$.
The RDF describes how particle density varies as a function of distance from a reference particle, revealing spatial organization and local structure in the system.
It is defined as:
\begin{equation}
    g(r) = \frac{1}{4 \pi r^2} \frac{1}{\rho n} \sum_i \sum_{j \neq i} \delta(r - \Vert \bm{x}_i - \bm{x}_j \Vert),
    \label{eq:rdf}
\end{equation}
where $\rho$ is the number density of atoms.
We average the RDF over the latter parts of the simulation, after discarding a short induction period (5 ps, 20\% of the trajectory).
The accuracy is quantified by the RDF MAE $= (1 / r_\text{cut}) \int_0^{r_\text{cut}} \vert \hat{g}(r) - g(r) \vert \, \mathrm{d}r$, with $r_\text{cut} = 5$ \AA{}.
Note that a similar set of metrics has been adopted in the benchmark of MLIP-based simulations \citep{fu2023forces}.

\section*{Data Availability}
The trajectories for universal MLIP dataset are available on Zenodo: \url{https://zenodo.org/doi/10.5281/zenodo.14889658} \cite{nam_2025_14889658}.
We obtained the LPS trajectories from \citet{jun2024nonexistence} directly from the authors and the LGPS trajectories from \citet{lopez2024how} are accessible at \url{https://superionic.upc.edu/}.
The script for preprocessing the training dataset from raw trajectories is available in the accompanying GitHub repository under the Code Availability section.

\section*{Code Availability}
The code to reproduce this work is available on GitHub: \url{https://github.com/learningmatter-mit/liflow} \cite{nam_2025_14889760}.

\begin{acknowledgments}
The authors thank Hoje Chun, Xiaochen Du, and MinGyu Choi for detailed feedback on the manuscript, Xiang Fu, Sungsoo Ahn, Johannes C. B. Dietschreit, and Mark Goldstein for their helpful suggestions and discussions, and Kacper Kapu\'sniak for providing the preliminary codebase for their work.
We acknowledge the MIT SuperCloud and Lincoln Laboratory Supercomputing Center for providing HPC resources.
J.N. acknowledge support from the Toyota Research Institute.
J.N. and K.J. were partially supported by the Energy Storage Research Alliance ``ESRA'' (DE-AC02-06CH11357), an Energy Innovation Hub funded by the U.S. Department of Energy, Office of Science, Basic Energy Sciences.
\end{acknowledgments}

\section*{Author Contributions}
J.N. and R.G-B. conceived the project.
J.N. and S.L. designed the machine learning methodologies and developed the algorithms.
J.N., G.W., and K.J. processed the simulation dataset and analyzed the results.
R.G-B. supervised the research.
All authors discussed the results and contributed to the final manuscript.

\section*{Competing Interests}
The authors declare no competing interests.

\bibliography{main}


\widetext
\clearpage

\setcounter{figure}{0}
\setcounter{table}{0}
\renewcommand{\figurename}{Extended Data Fig.}
\renewcommand{\tablename}{Extended Data Table}
\crefalias{figure}{extendedfigure}
\crefalias{table}{extendedtable}

\begin{figure}[!ht]
\includegraphics[width=0.8\textwidth]{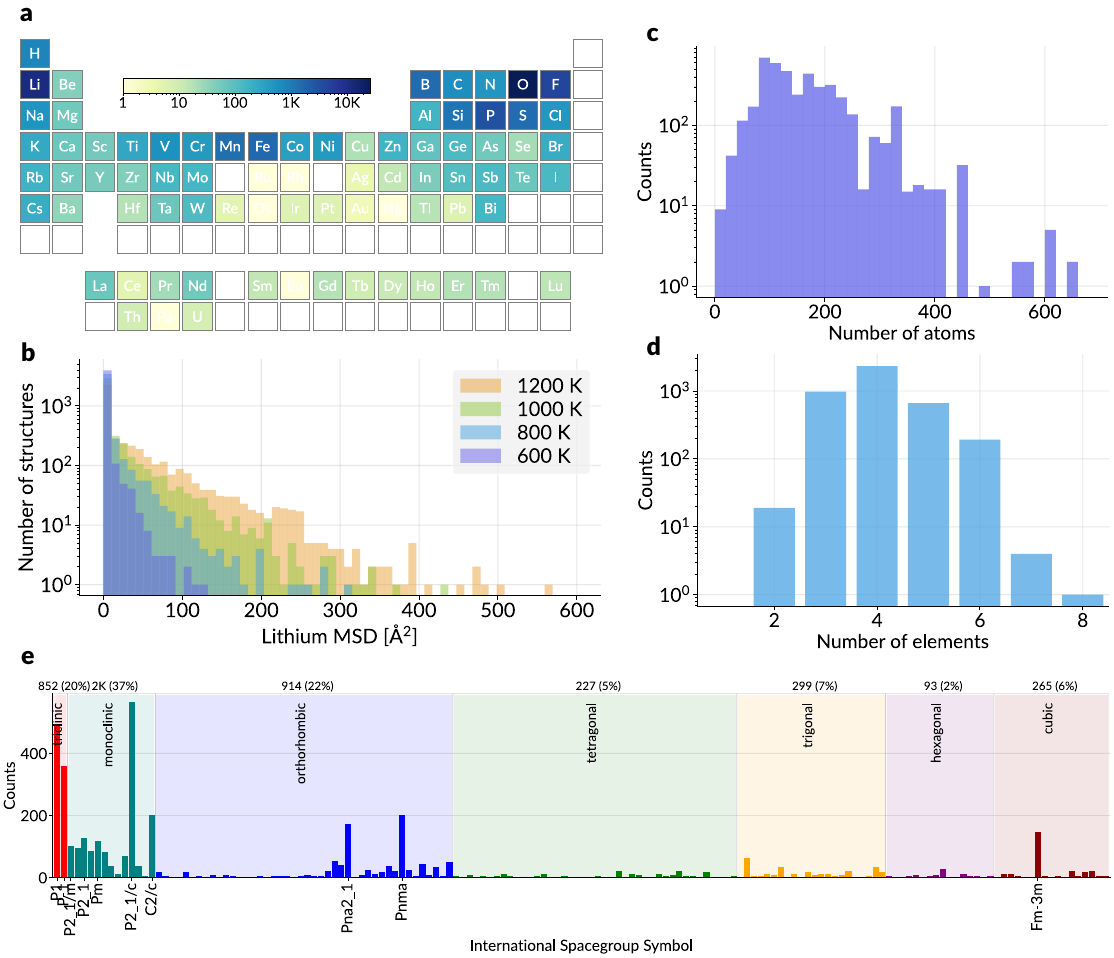}
\caption{
\textbf{Dataset statistics.}
\textbf{(a)} Elemental count distribution across the unit cells of the structures in the dataset.
\textbf{(b)} Histogram of lithium MSD values from 25-ps MD simulations at different temperatures.
\textbf{(c)} Distribution of atom counts (in the constructed supercell) per structure.
\textbf{(d)} Distribution of element counts per structure.
\textbf{(e)} Space group distribution of the structures.
}
\label{fig:data}
\end{figure}

\begin{figure}[!ht]
\includegraphics[height=0.9\textheight]{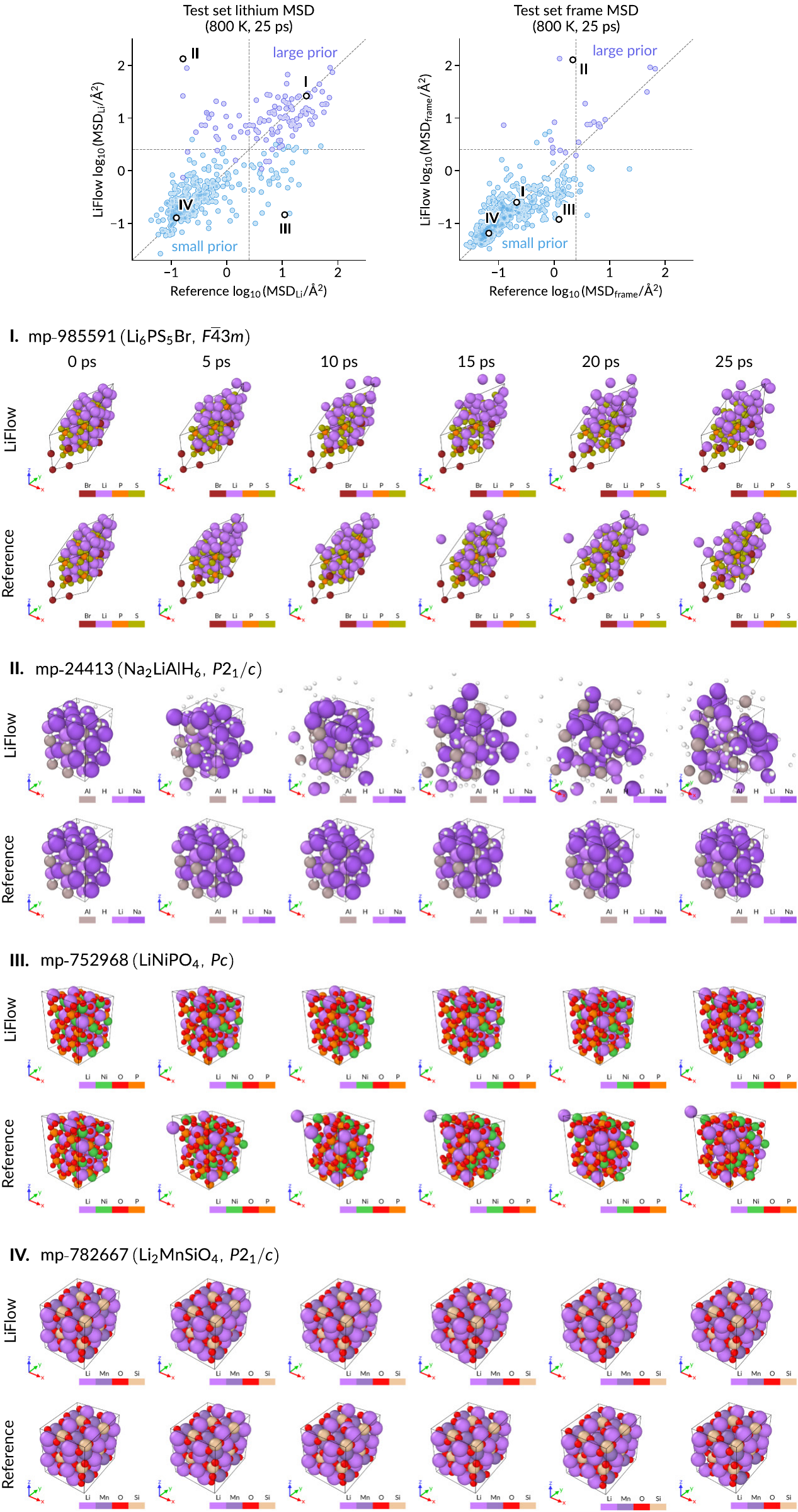}
\caption{
\textbf{Universal model inference example.}
(Top) Parity plots comparing the log MSD values for lithium and frame atoms in 800 K simulations (reference vs. 25-step LiFlow inference) across 419 test materials.
Data points are colored by their respective prior scales, with four annotated examples (\textbf{I}--\textbf{IV}) highlighted below.
\textbf{II} and \textbf{III} represent failed cases where lithium MSD is overestimated and underestimated, respectively.
Dotted lines indicate the classification boundary between large and small priors.
(Bottom) Reference and generated trajectories for the four annotated test set materials.
}
\label{fig:universal_example}
\end{figure}

\begin{figure}[!ht]
\includegraphics[width=0.45\textwidth]{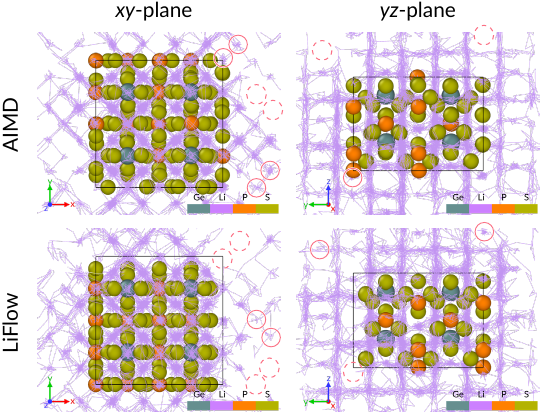}
\caption{
\textbf{Diffusion trace of lithium in LGPS simulations.}
The diffusion traces of lithium atoms for 150 ps trajectories using LiFlow and AIMD at 900 K.
Different lithium sites are accessed in different simulations, as indicated by circles: solid circles represent sites visited in the current simulation, while dotted circles indicate sites not visited in this simulation but visited in another.
}
\label{fig:diffusion_trace}
\end{figure}

\begin{figure}[!ht]
\includegraphics[width=0.7\textwidth]{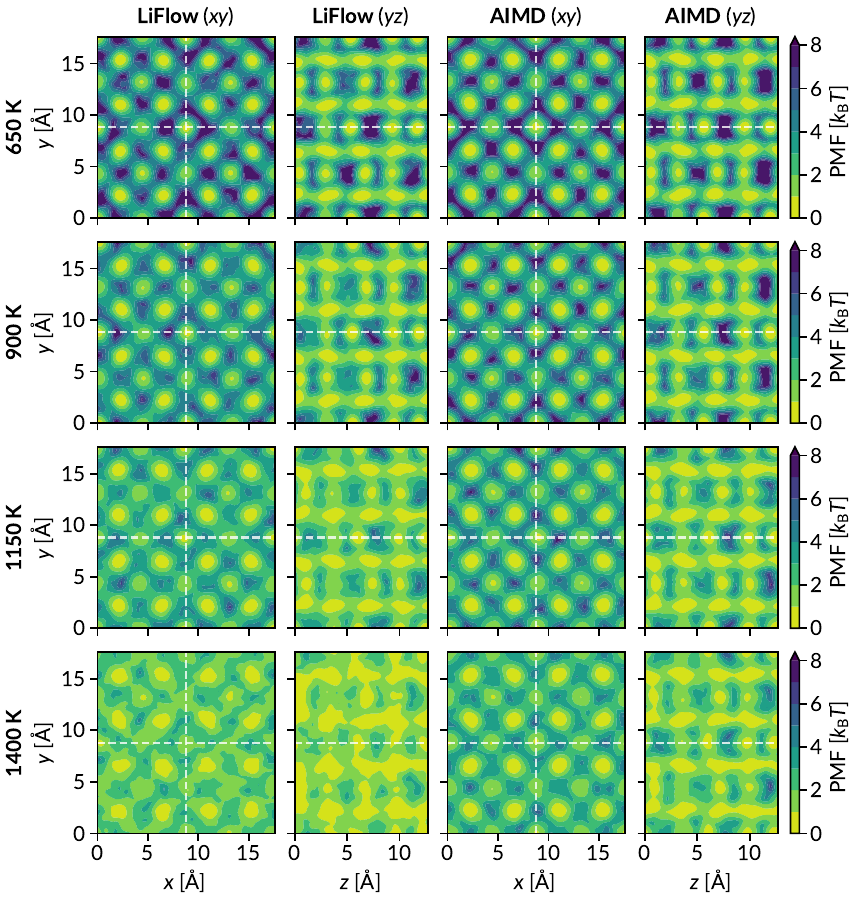}
\caption{
\textbf{Potentials of mean force for lithium in LGPS simulations.}
Potentials of mean force (PMFs) in units of $k_\text{B}T$ for lithium atoms in wrapped coordinates, shown for 150 ps trajectories using LiFlow and AIMD across different temperatures.
For each method, the first and second columns correspond to projections along the $x$--$y$ and $y$--$z$ planes, respectively.
Dotted lines indicate supercell boundaries.
}
\label{fig:pmf}
\end{figure}

\begin{figure}[!ht]
\begin{center}
\includegraphics[width=0.4\textwidth]{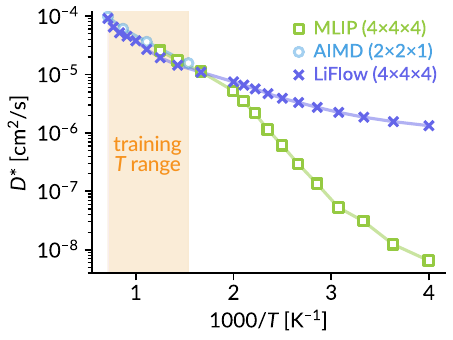}
\caption{
\textbf{Temperature extrapolation.}
Lithium self-diffusivity ($D^\ast$) plotted as a function of $1000/T$ for \ce{Li10GeP2S12} (LGPS), extending the data from \cref{fig:diffusion_AIMD}c to lower temperatures (higher $1000/T$).
}
\label{fig:diffusion_LGPS_ext}
\end{center}
\end{figure}

\begin{figure}[!ht]
\includegraphics[width=0.7\textwidth]{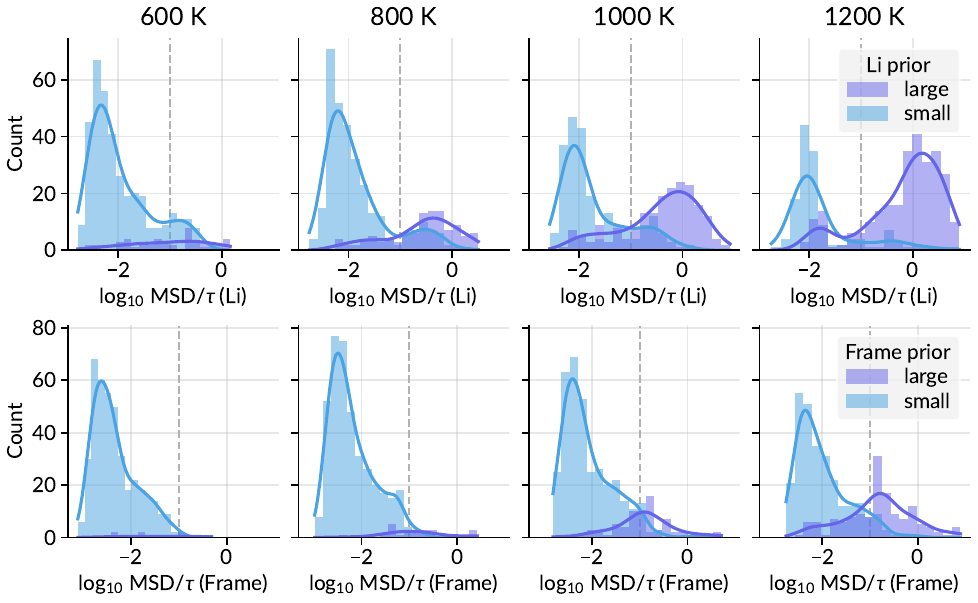}
\caption{
\textbf{Prior selector model performance.}
Histogram of the target values, $\log_{10} (\text{MSD}_\mathcal{S} / \tau)$, for lithium and frame atoms, colored by the predicted prior scale (small or large) for the test set materials.
The reference classification threshold ($-1.0$) is marked by a vertical dotted line.
$\text{MSD}/\tau$ values are reported in units of \AA{}$^2$/ps.
}
\label{fig:prior_scale_hist}
\end{figure}


\clearpage

\begin{center}
\textbf{\large Supplementary Note for: \\[1ex] Flow Matching for Accelerated Simulation of Atomic Transport in Materials}
\end{center}

\setcounter{equation}{0}
\setcounter{figure}{0}
\setcounter{table}{0}
\setcounter{section}{0}
\setcounter{page}{1}
\renewcommand{\theequation}{S\arabic{equation}}
\renewcommand{\thefigure}{S\arabic{figure}}
\renewcommand{\thetable}{S\arabic{table}}
\renewcommand{\thesection}{\Alph{section}}
\renewcommand{\thesubsection}{\arabic{subsection}}
\renewcommand{\figurename}{Fig.}
\renewcommand{\tablename}{Table}
\crefalias{figure}{suppfigure}
\crefalias{table}{supptable}
\crefalias{section}{suppsection}

{
\section*{Table of Contents}
\setstretch{2.0}
\contentsmargin{2.55em}
\dottedcontents{section}[3.8em]{}{2.3em}{1pc}
\dottedcontents{subsection}[7.6em]{}{2.3em}{1pc}

\startcontents
\printcontents{}{1}{}{}
}
\clearpage

\section{Proof for Proposition 1}
\label{si:proof_prop_1}

\setcounter{proposition}{0}
\begin{proposition}
    Given an invariant base distribution $p_0(\bm{D}_0)$ satisfying \cref{eq:sym_perm,eq:sym_rot} and an equivariant conditional vector field $u_t(\bm{D}_t \vert \bm{D}_1)$ with the following properties:
    \begin{align}
        u_t(\bm{P}\bm{D}_t \vert \bm{P}\bm{D}_1, \bm{P}\bm{X}, \bm{L}, \bm{P}\bm{a}, T) &= \bm{P} u_t(\bm{D}_t \vert \bm{D}_1, \bm{X}, \bm{L}, \bm{a}, T), &\qquad \bm{P} \in S_n \\
         u_t(\bm{D}_t\bm{R} \vert \bm{D}_1\bm{R}, \bm{X}\bm{R}, \bm{L}\bm{R}, \bm{a}, T) &= u_t(\bm{D}_t \vert \bm{D}_1, \bm{X}, \bm{L}, \bm{a}, T) \bm{R}, &\qquad \bm{R} \in \mathrm{O}(3)
    \end{align}
    the generated conditional probability path $p_{t \vert 1}(\bm{D}_t \vert \bm{D}_1)$ is invariant.
    Furthermore, given that the data distribution $q(\bm{D}_1)$ is invariant, the marginal probability path $p_t(\bm{D}_t)$ is also invariant.
\end{proposition}

\textit{Proof.}
We will prove for the $\mathrm{O}(3)$ symmetry, with a similar approach applying to $S_n$.
We omit the conditional variables ($\bm{X}, \bm{L}, \bm{a}, T$), as their transformations under group actions are implied by those of $\bm{D}_1$, either remaining invariant or transforming equivariantly.
The first part of the proof follows from Theorems 1 and 2 in \citet{kohler2020equivariant}, with additional conditional variables.
The conditional flow generated by the conditional vector field is
\begin{equation}
    \psi_t(\bm{D}_0 \vert \bm{D}_1) = \bm{D}_0 + \int_0^t u_s(\bm{D}_s \vert \bm{D}_1) \, \mathrm{d}s.
\end{equation}
Now, we apply $\bm{R} \in \mathrm{O}(3)$:
\begin{align}
    \psi_t(\bm{D}_0\bm{R} \vert \bm{D}_1\bm{R}) &= \bm{D}_0 \bm{R} + \int_0^t u_s(\bm{D}_s\bm{R} \vert \bm{D}_1\bm{R}) \, \mathrm{d}s \nonumber \\
    &= \bm{D}_0 \bm{R} + \int_0^t u_s(\bm{D}_s \vert \bm{D}_1) \bm{R} \, \mathrm{d}s \nonumber \\
    &= \left( \bm{D}_0 + \int_0^t u_s(\bm{D}_s \vert \bm{D}_1) \, \mathrm{d}s \right) \bm{R} \nonumber \\
    &= \psi_t(\bm{D}_0 \vert \bm{D}_1) \bm{R}.
\end{align}
Thus, the conditional flow $\psi_t$ is also equivariant with respect to $\bm{R}$.
Now, the conditional probability path $p_{t \vert 1}(\bm{D}_t \vert \bm{D}_1)$ is obtained as the pushforward of the prior distribution $p_0$ under $\psi_t$:
\begin{equation}
    p_{t \vert 1}(\bm{D}_t \vert \bm{D}_1) = [\psi_t]_{\#} p_0(\bm{D}_0) = p_0 \left( \psi_t^{-1}(\bm{D}_t \vert \bm{D}_1) \right) \left\vert \det \frac{\partial \psi_t^{-1}}{\partial \bm{D}_t} (\bm{D}_t \vert \bm{D}_1) \right\vert.
\end{equation}
Again, we apply $\bm{R} \in \mathrm{O}(3)$:
\begin{align}
    p_{t \vert 1}(\bm{D}_t\bm{R} \vert \bm{D}_1\bm{R}) &= p_0 \left( \psi_t^{-1}(\bm{D}_t\bm{R} \vert \bm{D}_1\bm{R}) \right) \left\vert \det \frac{\partial \psi_t^{-1}}{\partial (\bm{D}_t\bm{R})} (\bm{D}_t\bm{R} \vert \bm{D}_1\bm{R}) \right \vert \nonumber \\
    & = p_0 \left( \psi_t^{-1}(\bm{D}_t \vert \bm{D}_1) \bm{R} \right) \left\vert \det \frac{\partial \psi_t^{-1}}{\partial (\bm{D}_t\bm{R})} (\bm{D}_t\bm{R} \vert \bm{D}_1\bm{R}) \right \vert \nonumber \\
    & = p_0 \left( \psi_t^{-1}(\bm{D}_t \vert \bm{D}_1) \right) \left\vert \det \bm{I}_n \otimes \bm{R} \right\vert \left\vert \det \frac{\partial \psi_t^{-1}}{\partial \bm{D}_t} (\bm{D}_t \vert \bm{D}_1) \right\vert \left\vert \det \bm{I}_n \otimes \bm{R} \right\vert^{-1} \nonumber \\
    & = p_0 \left( \psi_t^{-1}(\bm{D}_t \vert \bm{D}_1) \right) \left\vert \det \frac{\partial \psi_t^{-1}}{\partial \bm{D}_t} (\bm{D}_t \vert \bm{D}_1) \right\vert \nonumber \\
    & = p_{t \vert 1}(\bm{D}_t \vert \bm{D}_1),
\end{align}
where we used the fact that $\left\vert \det \bm{I}_n \otimes \bm{R} \right\vert = \left\vert\det \bm{R} \right\vert^n = 1$.
Therefore, the resulting conditional probability path $p_{t \vert 1}$ is also invariant with respect to $\bm{R}$.

Now, for the marginal probability $p_t(\bm{D}_t) = \int p_{t \vert 1}(\bm{D}_t \vert \bm{D}_1) q(\bm{D}_1) \mathrm{d} \bm{D}_1$,
\begin{align}
    p_t(\bm{D}_t\bm{R}) &= \int p_{t \vert 1}(\bm{D}_t\bm{R} \vert \bm{D}_1\bm{R}) q(\bm{D}_1\bm{R}) \, \mathrm{d} (\bm{D}_1\bm{R}) \nonumber \\
    & = \int p_{t \vert 1}(\bm{D}_t \vert \bm{D}_1) q(\bm{D}_1) \, \left\vert \det \bm{I}_n \otimes \bm{R} \right\vert \mathrm{d} \bm{D}_1 \nonumber \\
    & = \int p_{t \vert 1}(\bm{D}_t \vert \bm{D}_1) q(\bm{D}_1) \, \mathrm{d} \bm{D}_1 \nonumber \\
    & = p_t(\bm{D}_t),
\end{align}
which concludes the proof of the invariance of the marginal $p_t$.
\hfill$\square$

\clearpage

\section{Training and Inference Details}
\label{si:training_and_inference_algorithms}

The training algorithms for the \textit{Propagator} and \textit{Corrector} are shown in \cref{alg:propagator_training,alg:corrector_training}, respectively.
When training on the universal dataset, material compositions are sampled uniformly by assigning a sampling weight inversely proportional to the number of materials in the training set with that specific composition.

\begin{algorithm2e}[!ht]
\caption{\textit{Propagator} Training}
\label{alg:propagator_training}
\DontPrintSemicolon
\KwIn{Dataset of time-separated material structures $\mathcal{D}$}
\KwOut{Optimized \textit{Propagator} parameter $\theta$}
\vspace{1ex}
\While{Training}{
    Sample data $(\bm{X}_\tau, \bm{X}_{\tau + \Delta \tau}, \bm{L}, \bm{a}, T) \sim \mathcal{D}$ \;
    Sample flow time $t \sim \mathcal{U}(t; 0, 1)$ \;
    Sample \textit{Propagator} prior  $\bm{D}_0 \sim \mathcal{N}(\bm{D}_0; \bm{0}, \operatorname{diag}(\bm{\sigma})^2 \otimes \bm{I}_3)$ \;
    $\bm{D}_1 \gets \bm{X}_{\tau + \Delta \tau}$ \tcp*[r]{True displacements}
    $\bm{D}_t \gets (1 - t) \bm{D}_0 + t \bm{D}_1$ \tcp*[r]{Interpolated displacements (\cref{eq:ot_flow})}
    $u_t(\bm{D}_t \vert \bm{D}_1) \gets (\bm{D}_1 - \bm{D}_t) / (1 - t)$ \tcp*[r]{Conditional flow (\cref{eq:ot_flow})}
    $v_t(\bm{D}_t; \theta) \gets \text{\textit{Propagator}}(\bm{D}_t, \bm{X}_\tau, \bm{L}, \bm{a}, T, t ; \theta)$ \;
    $\mathcal{L}_\text{CFM}(\theta) \gets \Vert v_t(\bm{D}_t; \theta) - u_t(\bm{D}_t \vert \bm{D}_1) \Vert^2$ \tcp*[r]{CFM regression objective (\cref{eq:cfm_loss})}
    $\theta \gets \text{Update}(\theta, \nabla_\theta \mathcal{L}_\text{CFM}(\theta))$ \tcp*[r]{Parameter update}
}
\end{algorithm2e}

\begin{algorithm2e}[!ht]
\caption{\textit{Corrector} Training}
\label{alg:corrector_training}
\DontPrintSemicolon
\KwIn{Dataset of time-separated material structures $\mathcal{D}$}
\KwOut{Optimized \textit{Corrector} parameter $\theta$}
\vspace{1ex}
\While{Training}{
    Sample data $(\cdot, \bm{X}_{\tau}, \bm{L}, \bm{a}, T) \sim \mathcal{D}$ \;
    Sample flow time $t \sim \mathcal{U}(t; 0, 1)$ \;
    Sample \textit{Corrector} prior  $\bm{D}_0 \sim \mathcal{N}(\bm{D}_0; \bm{0}, \operatorname{diag}(\bm{\sigma})^2 \otimes \bm{I}_3)$ \;
    Sample noise scale $\bm{\sigma}' \sim \mathcal{U}(\bm{\sigma}'; \bm{0}, \sigma_\text{max} \bm{1}_n)$ \;
    Sample positional noise displacement $\bm{D} \vert \bm{\sigma}' \sim \mathcal{N}(\bm{D}; \bm{0}, \operatorname{diag}(\bm{\sigma}')^2 \otimes \bm{I}_3$) \;
    $\tilde{\bm{X}}_\tau \gets \bm{X}_\tau + \bm{D}$ \tcp*[r]{Noisy positions}
    $\bm{D}_1 \gets -\bm{D}$ \tcp*[r]{True denoising displacements}
    $\bm{D}_t \gets (1 - t) \bm{D}_0 + t \bm{D}_1$ \tcp*[r]{Interpolated displacements (\cref{eq:ot_flow})}
    $u_t(\bm{D}_t \vert \bm{D}_1) \gets (\bm{D}_1 - \bm{D}_t) / (1 - t)$ \tcp*[r]{Conditional flow (\cref{eq:ot_flow})}
    $v_t(\bm{D}_t; \theta) \gets \text{\textit{Corrector}}(\bm{D}_t, \tilde{\bm{X}}_\tau, \bm{L}, \bm{a}, T, t ; \theta)$ \;
    $\mathcal{L}_\text{CFM}(\theta) \gets \Vert v_t(\bm{D}_t; \theta) - u_t(\bm{D}_t \vert \bm{D}_1) \Vert^2$ \tcp*[r]{CFM regression objective (\cref{eq:cfm_loss})}
    $\theta \gets \text{Update}(\theta, \nabla_\theta \mathcal{L}_\text{CFM}(\theta))$ \tcp*[r]{Parameter update}
}
\end{algorithm2e}

\begin{algorithm2e}[!ht]
\caption{LiFlow Inference}
\label{alg:inference}
\DontPrintSemicolon
\KwIn{Initial position $\bm{X}_0$, lattice $\bm{L}$, atom types $\bm{a}$, atomic masses $\bm{m}(\bm{a})$, temperature $T$}
\KwOut{Predicted position $\bm{X}_{\tau}$ at $\tau = N_\text{step}\Delta \tau$}
\vspace{0.5ex}
Determine the prior from $\bm{X}_0$, $\bm{L}$, $\bm{a}$, $\bm{m}$, and $T$ \;
\For{$i_\tau \gets 0$ \KwTo $N_\textup{step} - 1$}{
    $\tau \gets i_\tau \Delta \tau$ and $\tau' \gets (i_\tau + 1) \Delta \tau$ \;
    Sample $\bm{D}$ from the \textit{Propagator} prior \;
    \For{$i \gets 0$ \KwTo $N_\textup{flow} - 1$}{
        $\bm{D} \gets \bm{D} + \text{\textit{Propagator}}(\bm{D}, \bm{X}_\tau, \bm{L}, \bm{a}, T, t) / N_\text{flow}$
    }
    $\tilde{\bm{X}}_{\tau'} \gets \bm{X}_{\tau} + \bm{D}$ {\tcp*[r]{\textit{Propagator} step}}
    Sample $\bm{D}$ from the \textit{Corrector} prior \;
    \For{$i \gets 0$ \KwTo $N_\textup{flow} - 1$}{
        $\bm{D} \gets \bm{D} + \text{\textit{Corrector}}(\bm{D}, \tilde{\bm{X}}_{\tau'}, \bm{L}, \bm{a}, T, t) / N_\text{flow}$
    }
    $\bm{X}_{\tau'} \gets \tilde{\bm{X}}_{\tau'} + \bm{D}$ {\tcp*[r]{\textit{Corrector} step}}
    $\bm{X}_{\tau'} \gets \bm{X}_{\tau'} - \text{CoM}(\bm{X}_{\tau'}, \bm{m}) + \text{CoM}(\bm{X}_\tau, \bm{m})$ \;
}
\end{algorithm2e}

Training hyperparameters are provided in \cref{table:hyperparameter}.
Additional results on prior scale and PaiNN model ablations are shown in \cref{table:prior_iso_maxwell} and \cref{table:painn_ablation}, respectively.

\begin{table}[!ht]
\caption{
\textbf{Training hyperparameters.}
Hyperparameters for training the \textit{Propagator} and \textit{Corrector} models.
}
\label{table:hyperparameter}
\begin{tblr}{colspec=ll,rowsep=1pt,colsep=8pt,vline{2}}
\toprule
Parameter & Value \\
\midrule
Feature dimension & 64 \\
Radial basis functions & 20 \\
Message passing layers & 3 \\
Cutoff distance (\AA) & 5.0 \\
Offset distance (\AA) & 0.5 \\
Optimizer & Adam \\ 
Learning rate & 0.0003 \\
Gradient clipping norm & 10.0 \\
Batch size & 16 \\
Maximum training steps & 125,000 \\
\bottomrule
\end{tblr}
\end{table}

\begin{table}[!ht]
\caption{
\textbf{Effect of prior design.}
Comparison between the isotropic prior and the scaled Maxwell--Boltzmann prior using different scale multipliers.
Only the \textit{Propagator} model was used, and trained and tested on 800 K trajectories.
Standard deviations (in parentheses) are from three independent runs.
MSD values are reported in units of \AA{}$^2$, the best overall metric values are shown in bold, and the best for each prior are underlined.
}
\label{table:prior_iso_maxwell}
\begin{tblr}{
    colspec=clcccc,rowsep=1pt,colsep=8pt,vline{3},
    cell{2}{1}={r=5}{valign=m},cell{7}{1}={r=5}{valign=m} 
}
\toprule
Prior & \makecell{Scale \\ multiplier ($\sigma$)} & \makecell{log MSD\textsubscript{Li} \\ MAE ($\downarrow$)} & \makecell{log MSD\textsubscript{Li} \\ $\rho$ ($\uparrow$)} & \makecell{log MSD\textsubscript{frame} \\ MAE ($\downarrow$)} & \makecell{Stable traj. \\ \% ($\uparrow$)} \\
\midrule
Isotropic & $10^{-2}$ & 0.726 (0.012) & 0.550 (0.008) & 0.900 (0.007) & 99.9 (0.1) \\
&  $10^{-1.5}$ &  \underline{0.498} (0.003) &  \underline{0.753} (0.008) &  \underline{0.318} (0.008) &  98.6 (0.2) \\
& $10^{-1}$ & 0.531 (0.008) & 0.713 (0.008) & 0.454 (0.012) & 95.9 (0.2) \\
& $10^{-0.5}$ & 0.551 (0.009) & 0.723 (0.004) & 0.470 (0.001) & \textbf{\underline{100.0}} (0.0) \\
& $10^{0}$ & 0.626 (0.005) & 0.712 (0.004) & 0.408 (0.002) & \textbf{\underline{100.0}} (0.0) \\
\midrule
\makecell{Maxwell-- \\ Boltzmann} & $10^{-1}$ & 0.694 (0.002) & 0.563 (0.007) & 0.653 (0.003) & 88.1 (1.5) \\
& $10^{-0.5}$ & 0.511 (0.004) & 0.682 (0.006) & 0.419 (0.004) & 99.8 (0.2) \\
&  $10^{0}$ &  \textbf{\underline{0.396}} (0.006) &  \textbf{\underline{0.779}} (0.009) &  \textbf{\underline{0.274}} (0.003) &  99.4 (0.1) \\
& $10^{0.5}$ & 0.654 (0.002) & 0.694 (0.006) & 0.447 (0.005) & 99.4 (0.1) \\
& $10^{1}$ & 0.577 (0.007) & 0.709 (0.009) & 0.339 (0.007) & \underline{99.9} (0.1) \\
\bottomrule
\end{tblr}
\end{table}

\begin{table}[!ht]
\caption{
\textbf{Effect of PaiNN modification.}
Evaluation metrics for the \textit{Propagator} model using a uniform scale ($\sigma = 1$) Maxwell--Boltzmann prior distribution.
Standard deviations (in parentheses) are from three independent runs.
MSD values are reported in units of \AA{}$^2$, and the best metric values are shown in bold.
}
\label{table:painn_ablation}
\begin{tblr}{
    colspec=ccccccc,rowsep=1pt,colsep=8pt,vline{4},
    cell{2}{1}={r=2}{valign=m},cell{2}{2}={r=2}{valign=m} 
}
\toprule
\makecell{Train \\ $T$ (K)} & \makecell{Inference \\ $T$ (K)} & \makecell{Model} & \makecell{log MSD\textsubscript{Li} \\ MAE ($\downarrow$)} & \makecell{log MSD\textsubscript{Li} \\ $\rho$ ($\uparrow$)} & \makecell{log MSD\textsubscript{frame} \\ MAE ($\downarrow$)} & \makecell{Stable traj. \\ \% ($\uparrow$)} \\
\midrule
800 & 800 & PaiNN & 0.976 (0.008) & 0.344 (0.005) & 1.217 (0.009) & 38.1 (1.3) \\
&& Modified PaiNN & \textbf{0.396} (0.006) & \textbf{0.779} (0.009) & \textbf{0.274} (0.003) & \textbf{99.4} (0.1) \\
\bottomrule
\end{tblr}
\end{table}

\clearpage
\section{Equivariance ablation}
\label{si:equivariance_ablation}

Furthermore, we conducted an ablation study replacing the PaiNN \citep{schutt2021equivariant} backbone in the LiFlow model with the GNS \citep{sanchezconzalez2020learning} backbone used in Orb \citep{neumann2024orb}, which is a state-of-the-art non-equivariant MLIP.
We matched the model hyperparameters to maintain a similar total parameter count (LiFlow: 174k, LiFlow-GNS: 206k), applied rotational augmentation to the input structure pairs, and trained the LiFlow-GNS model using identical settings aside from the backbone architecture.

\begin{figure}[!ht]
\centering
\includegraphics[width=0.5\textwidth]{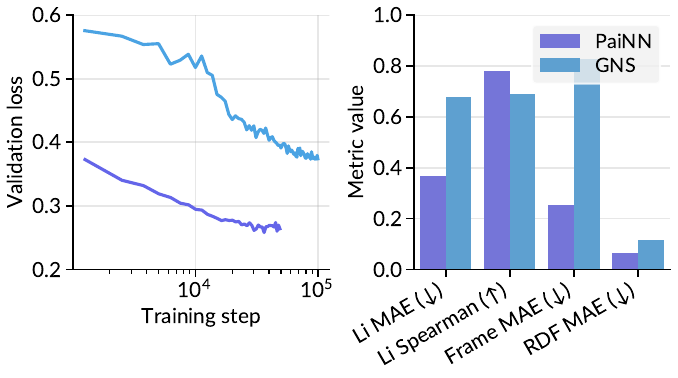}
\caption{
\textbf{Equivariance ablation.}
(Left) Validation loss during LiFlow \textit{Propagator} training using PaiNN (equivariant) and GNS (non-equivariant) as model backbones.
(Right) Result metrics from the main text for 800 K inference on the universal MLIP dataset using both backbones.
``Li'' and ``Frame'' in the labels indicate log(MSD/\AA{}$^2$) values for each species, respectively.
}
\label{fig:ablation_equiv}
\end{figure}

The results in \cref{fig:ablation_equiv} show that the non-equivariant architecture yields higher validation loss and consequently inferior metric values.
Although this is not a comprehensive comparison of equivariant and non-equivariant architectures or training schemes, the findings suggest that, for the given dataset size and generative modeling task, incorporating equivariance into the model architecture leads to more efficient use of the available simulation data.

\clearpage
\section{Batched inference}
\label{si:batched_inference}

\begin{figure*}[!ht]
\centering
\includegraphics[width=0.5\textwidth]{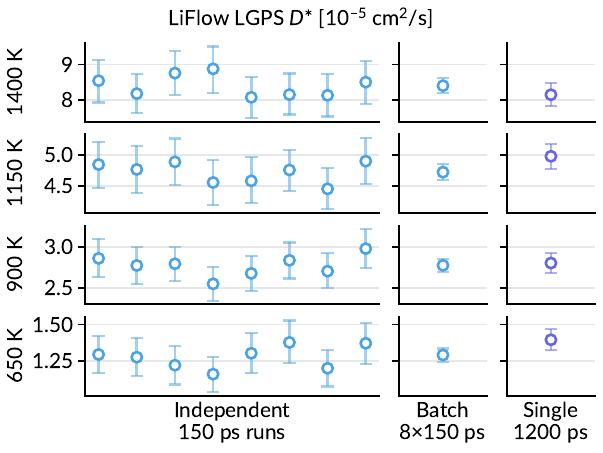}
\caption{
\textbf{Batched evaluation.}
The self-diffusivity $D^\ast$ values obtained from LiFlow trajectories at respective temperatures.
(Left) $D^\ast$ obtained from batched eight independent 150 ps trajectories with different initial seeds.
(Middle) $D^\ast$ obtained by collecting all MSD from eight independent 150 ps trajectories in the batch.
(Right) $D^\ast$ obtained from a single 1,200 ps trajectory.
Scatter points and error bars represent the median and 95\% confidence intervals (CIs) for $D^\ast$, from 1,000 MCMC samples of the Bayesian regression \citep{mccluskey2024kinisi}.
}
\label{fig:batched_eval}
\end{figure*}

For the LiFlow model trained on AIMD trajectories of \ce{Li10GeP2S12} (LGPS), which was used for the results in \cref{fig:diffusion_AIMD}b and c of the main text, we ran batched short simulations of 150 ps with eight different seeds in parallel and a single long simulation of 1,200 ps at each temperature (650, 900, 1150, 1400 K).
The resulting self-diffusivities, obtained as detailed in the Methods section, are reported in \cref{fig:batched_eval}.

From \cref{fig:batched_eval}, we observe that different runs yield consistent $D^\ast$ values at each temperature.
$D^\ast$ from the collected batch and the single 1,200 ps trajectory exhibit smaller CI ranges than individual runs.
The 1,200 ps trajectory has a relatively wider CI range than the aggregated batch.
This is because MSD windows in a single trajectory are more temporally correlated.

\clearpage

\section{Hyperparameter Sensitivity}
\label{si:hyperparameter_sensitivity}

\begin{figure}[!ht]
\includegraphics[width=0.8\textwidth]{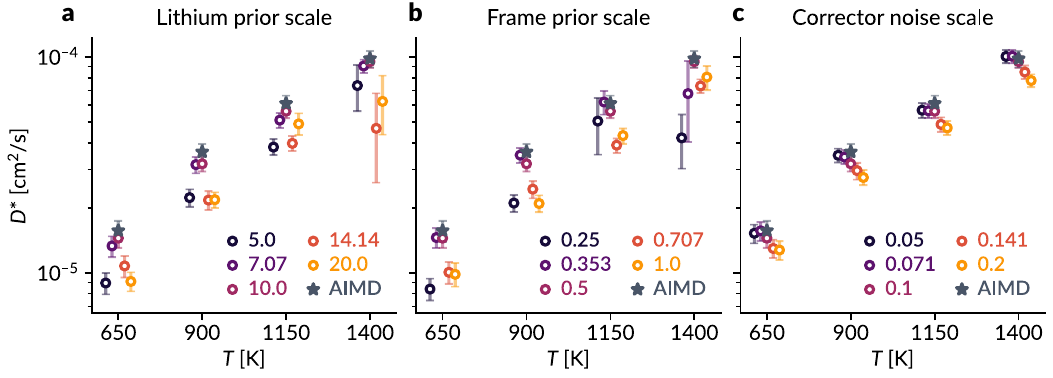}
\caption{
\textbf{Scale hyperparameter sensitivity for LGPS models.}
\textbf{(a)} Variation of the \textit{Propagator} lithium prior scale (default: 10.0).
\textbf{(b)} Variation of the \textit{Propagator} frame prior scale (default: 0.5).
\textbf{(c)} Variation of the \textit{Corrector} noise scale (default: 0.1).
Results from AIMD reference simulations are also included.
Scatter points and error bars represent the median and 95\% confidence intervals (CIs) for $D^\ast$, from 1,000 MCMC samples of the Bayesian regression \citep{mccluskey2024kinisi}.
Prior scales are given in units of $\text{\AA{}}\cdot (\text{eV} \cdot \text{K} / \text{u})^{-1/2}$ (see Methods section in the main text).
}
\label{fig:scale_ablation}
\end{figure}

To evaluate the impact of prior and noise distribution scale hyperparameters on predicting kinetic properties, we perform a sensitivity analysis using the LGPS dataset.
For the \textit{Propagator} scales (lithium and frame) and the \textit{Corrector} noise scale, we vary the scales from $\times 1/2$ to $\times 2$, train the corresponding models, and conduct a 150-step (150.75 ps) LiFlow inference for each model as described in the main text.
Results in \cref{fig:scale_ablation} demonstrate that diffusivity values show minor deviations from their peak value at the optimal \textit{Propagator} prior scales.
Changing \textit{Corrector} noise scale as in \cref{fig:scale_ablation}c demonstrates that the \textit{Corrector} noise scale larger than a certain threshold causes diffusivities to decrease, suggesting that stronger correction enhances stability but diminishes diffusive behavior slightly.

\begin{figure}[!ht]
\includegraphics[width=0.8\textwidth]{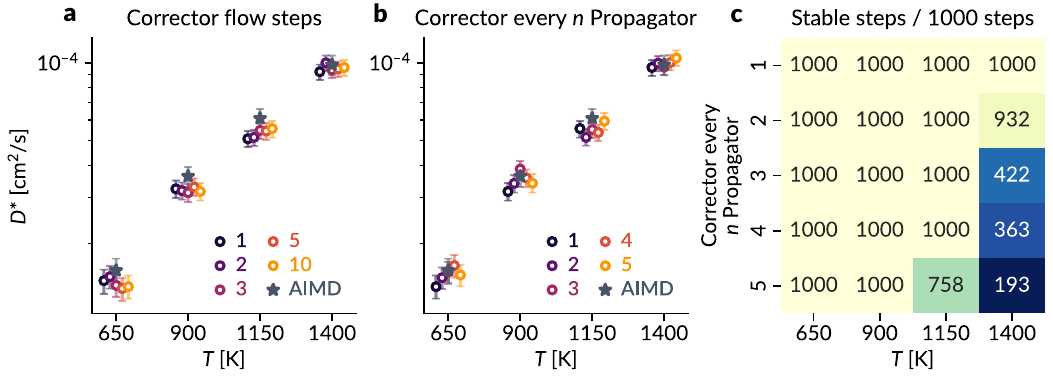}
\caption{
\textbf{\textit{Corrector} inference ablation for LGPS models.}
\textbf{(a)} Variation of the \textit{Corrector} flow steps ($N_\text{flow}$, default: 10).
\textbf{(b)} Applying the \textit{Corrector} every $n$ \textit{Propagator} steps (default: 1).
Results from AIMD reference simulations are also included.
Scatter points and error bars represent the median and 95\% confidence intervals (CIs) for $D^\ast$, from 1,000 MCMC samples of the Bayesian regression \citep{mccluskey2024kinisi}.
\textbf{(c)} Number of stable propagation steps over a 1,000-step inference.
}
\label{fig:corrector_ablation}
\end{figure}

While the \textit{Corrector} significantly improves inference for materials with varying compositions (universal dataset, \cref{table:universal_result}), we found that it plays a reduced role in AIMD models, where training and inference involve the same material structure, as the \textit{Propagator} is sufficiently trained to allow simplified \textit{Corrector} inference.
In \cref{fig:corrector_ablation}a and b, we analyze reducing \textit{Corrector} flow steps and performing \textit{Corrector} inference every $n$ \textit{Propagator} steps (e.g., \textit{PPPCPPPC}$\cdots$ for $n = 3$ versus \textit{PCPCPC}$\cdots$ for $n = 1$).
Diffusivity values remain largely unaffected in both cases.
However, when we extend the inference to 1,000 steps (1.005 ns, \cref{fig:corrector_ablation}c), we could observe that higher $n$ values lead to propagation instability at elevated temperatures.

\clearpage

\section{Dataset Scaling}
\label{si:dataset_scaling}

\begin{figure}[!ht]
\centering
\includegraphics[width=0.7\textwidth]{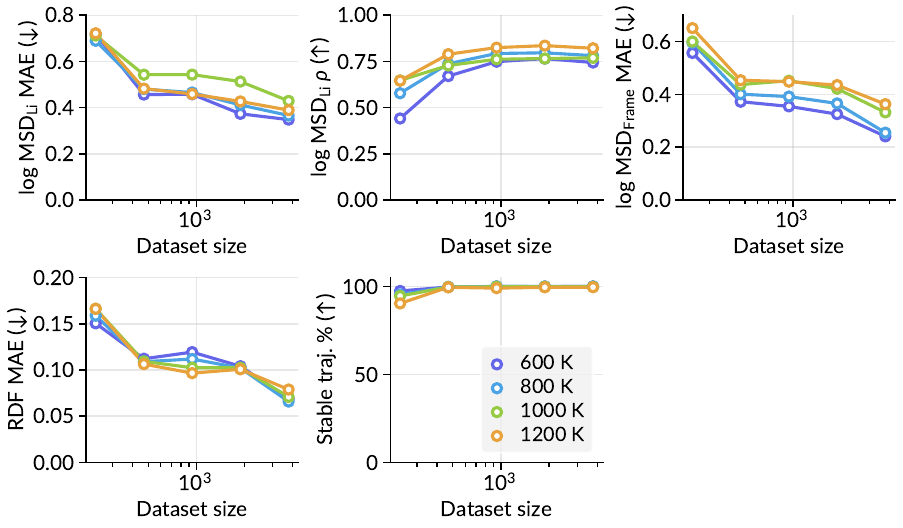}
\caption{
\textbf{Training dataset scaling.}
Result metrics for 25-ps inference on test set materials at each temperature using LiFlow models trained on subsets of the universal MLIP dataset.
Training set sizes were progressively halved by randomly sampling from the original 3,767 trajectories.
$\rho$ denotes the Spearman correlation, and MSD values are reported in units of \AA{}$^2$.
}
\label{fig:ablation_dataset}
\end{figure}

Here, we present scaling curves that show how result metrics vary with training set size for the universal MLIP dataset.
Starting from the full dataset of 3,767 trajectories, we progressively halved the size by randomly subsampling materials.
Using consistent hyperparameters for both the \textit{Propagator} and \textit{Corrector}, we performed 25-ps inference on test set materials and compared the predictions to reference trajectories.
The result metrics, as described in the main text, are shown in \cref{fig:ablation_dataset}.

Trajectory accuracy, measured by log MSD MAE and RDF MAE, improves steadily with increasing dataset size, suggesting room for further gains.
In contrast, the Spearman correlation ($\rho$) of lithium log MSD nearly saturates, indicating that relative differences between materials are easier to learn than precise dynamical observables.

\clearpage

\clearpage

\section{Out-of-distribution Inference}
\label{si:ood_evaluation}

\begin{figure}[!ht]
\centering
\includegraphics[width=0.65\textwidth]{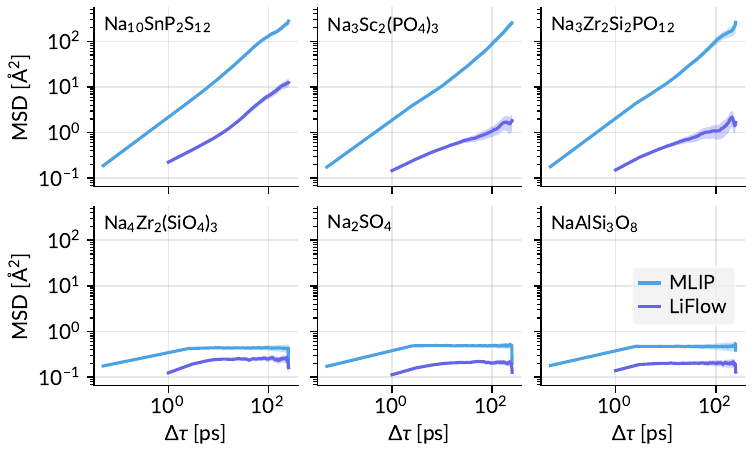}
\caption{
\textbf{Out-of-distribution evaluation of sodium-containing materials.}
Time evolution of mean-squared displacement (MSD) over 250-ps trajectories from MLIP simulations (MACE-MP-0 small model \citep{batatia2024foundation}) and zero-shot LiFlow inference at 800 K.
Top row: good sodium-ion conductors (\ce{Na10SnP2S12}, \ce{Na3Sc2(PO4)3}, and \ce{Na3Zr2Si2PO12}).
Bottom row: poor or non-conductors (\ce{Na4Zr2(SiO4)3}, \ce{Na2SO4}, and \ce{NaAlSi3O8}).
Solid lines indicate the mean,W and shaded regions indicate standard deviation from bootstrapping time windows within each trajectory.
}
\label{fig:ood_msd_eval}
\end{figure}

While the methods and evaluations in the main text primarily focus on lithium-based SSEs, alternative mobile-ion chemistries, such as those based on sodium or zinc, are also of growing interest.
With appropriate adjustments to sampling intervals and prior scale selection, the presented method can be extended to these systems.
Here, we assess the preliminary generalization capability of the current LiFlow model to out-of-distribution (OOD) sodium-containing materials without additional fine-tuning.
We selected three good sodium-ion conductors: \ce{Na10SnP2S12}, \ce{Na3Sc2(PO4)3}, and \ce{Na3Zr2Si2PO12}, as well as three poor or non-conductors: \ce{Na4Zr2(SiO4)3}, \ce{Na2SO4}, and \ce{NaAlSi3O8}, to evaluate and compare their diffusive behavior.
For each material, we performed NVT MD simulations at 800 K using the MACE-MP-0 small model \citep{batatia2024foundation}, following the same procedure used to construct the universal MLIP dataset.
We also ran zero-shot LiFlow inference at 800 K.
Due to the slower diffusion of Na ions, both trajectories were extended to 250 ps.
The time evolution of MSD for each material is shown in \cref{fig:ood_msd_eval}.

While MSD values are underestimated by one to two orders of magnitude for sodium conductors, the LiFlow model successfully distinguishes conductors from non-conductors by generating diffusive behavior for sodium ions in conductive materials.
Despite being trained predominantly on lithium-based materials, the model qualitatively captures ion transport in OOD sodium systems.
This suggests that the generative model encodes shared representations across related chemistries, which could be leveraged through fine-tuning for specific ion types or by developing a larger-scale universal model adaptable to diverse chemistries.

\clearpage
\section{Correlated Transport}
\label{si:correlated_transport}

\begin{figure}[!ht]
\centering
\includegraphics[width=0.7\textwidth]{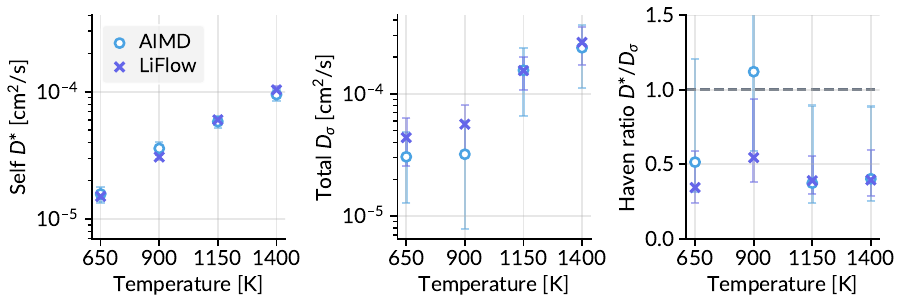}
\caption{
\textbf{Correlated transport metrics.}
Transport metrics computed from the 150 ps LGPS trajectories (AIMD reference and LiFlow) at 650, 900, 1150, and 1400 K used in Fig.~3b of the main text.
(Left) Lithium self-diffusivity $D^\ast$ (\cref{eq:si_self_diffusivity}).
(Center) Lithium total diffusivity $D_\sigma$ (\cref{eq:si_total_diffusivity}).
(Right) Haven ratio $H_R = D^\ast / D_\sigma$.
Scatter points and error bars show the median and 95\% confidence intervals (CIs) from 1,000 MCMC samples of the Bayesian regression \citep{mccluskey2024kinisi}.
}
\label{fig:haven_ratio_LGPS}
\end{figure}

The main kinetic observables reported throughout the main text are the MSD and self-diffusivity $D^\ast$.
However, to fully characterize ionic conductivity, correlated transport must also be considered, especially in concentrated or highly correlated ion lattices.
To quantify this, in addition to $D^\ast$ (\cref{eq:si_self_diffusivity}), we compute the total diffusivity $D_\sigma$ (\cref{eq:si_total_diffusivity}), which are defined for species $\mathcal{S}$ as follows:
\begin{align}
    D^\ast &= \lim_{\tau \to \infty} \frac{1}{6\tau |\mathcal{S}|} \sum_{i \in \mathcal{S}} \Vert \bm{x}_{\tau, i} - \bm{x}_{0, i} \Vert^2, \label{eq:si_self_diffusivity} \\
    D_{\sigma} &= \lim_{\tau \to \infty} \frac{1}{6 \tau |\mathcal{S}|} \left\Vert \sum_{i \in \mathcal{S}}(\bm{x}_{\tau, i} - \bm{x}_{0, i}) \right\Vert^2. \label{eq:si_total_diffusivity}
\end{align}
As with $D^\ast$, $D_\sigma$ is computed by regressing total displacements from multiple time origins against the time intervals.
We use Bayesian regression as implemented in the \texttt{kinisi} package \citep{mccluskey2024accurate,mccluskey2024kinisi} to estimate diffusivities.
To quantify the degree of correlated motion, we use the Haven ratio $H_R$, defined as
\begin{equation}
    H_R = \frac{D^\ast}{D_\sigma}.
    \label{eq:si_haven_ratio}
\end{equation}
A value of $H_R < 1$ indicates correlated ion motion, while $H_R = 1$ corresponds to ideal, uncorrelated (tracer-like) diffusion.
Using the 150 ps LGPS trajectories from Fig.~3b in the main text, we additionally computed $D_\sigma$ and $H_R$ for lithium ions, as shown in \cref{fig:haven_ratio_LGPS}.

In \cref{fig:haven_ratio_LGPS}, the total diffusivity $D_\sigma$ and Haven ratio $H_R$ predicted by LiFlow closely match the reference simulations, with $H_R$ values around 0.3--0.4.
This is consistent with previous findings \citep{winter2023simulations} and indicates that lithium ion transport in LGPS at elevated temperatures is enhanced by electrostatic repulsion between neighboring ions, leading to correlated motion.
The median $H_R$ from the reference trajectory exceeds 1 at 900 K, likely due to underestimation of $D_\sigma$ from the limited 150 ps trajectory length.
Reliable estimation of $D_\sigma$ generally requires longer trajectories than $D^\ast$, as reflected in the wider confidence intervals.

\end{document}